\documentclass[aps,showpacs]{revtex4} 

\usepackage{amssymb} 
\usepackage{graphicx} 
\usepackage{bm} 
\usepackage[T1]{fontenc} 
\usepackage{euscript} 
\usepackage{mathrsfs}

\begin{document} 

\newcommand{\cequ}{\begin{eqnarray}} 
\newcommand{\fequ}{\end{eqnarray}} 
\newcommand{\anticomut}[2]{\left\{#1,#2\right\}} 
\newcommand{\comut}[2]{\left[#1,#2\right]} 
\newcommand{\comutd}[2]{\left[#1,#2\right]^{*}} 

\title{\bf 
Hamiltonian thermodynamics of three-dimensional dilatonic black holes} 
\author{Gon\c{c}alo A. S. Dias\footnote{Email: gadias@fisica.ist.utl.pt}} 
\affiliation{Centro Multidisciplinar de Astrof\'{\i}sica - CENTRA \\ 
Departamento de F\'{\i}sica, Instituto Superior T\'ecnico - IST,\\ 
Universidade T\'ecnica de Lisboa - UTL,\\ 
Avenida Rovisco Pais 1, 1049-001 Lisboa, Portugal} 
\author{Jos\'e P. S. Lemos\footnote{Email: lemos@fisica.ist.utl.pt}} 
\affiliation{Centro Multidisciplinar de {}Astrof\'{\i}sica - CENTRA \\ 
Departamento de F\'{\i}sica, Instituto Superior T\'ecnico - IST,\\ 
Universidade T\'ecnica de Lisboa - UTL,\\ 
Avenida Rovisco Pais 1, 1049-001 Lisboa, Portugal} 

\begin{abstract} 

The action for a class of three-dimensional dilaton-gravity theories
with a negative cosmological constant can be recast in a Brans-Dicke type
action, with its free $\omega$ parameter.  These theories have static
spherically symmetric black holes. Those with well formulated
asymptotics are studied through a Hamiltonian formalism, and their
thermodynamical properties are found out. The theories studied are
general relativity ($\omega\to\infty$), a dimensionally reduced
cylindrical four-dimensional general relativity theory ($\omega=0$),
and a theory representing a class of theories ($\omega=-3$).  The
Hamiltonian formalism is setup in three dimensions through 
foliations on the right region of the
Carter-Penrose diagram, with the bifurcation 1-sphere as the left
boundary, and anti-de Sitter infinity as the right boundary.  The
metric functions on the foliated hypersurfaces are the canonical
coordinates. The Hamiltonian action is written, the Hamiltonian being
a sum of constraints.  
One finds a  new action which yields 
an unconstrained theory with one pair of canonical
coordinates $\{M,P_M\}$, $M$ being 
the mass parameter and
$P_M$ its conjugate momenta
The resulting Hamiltonian is a sum of boundary
terms only.  A quantization of the theory is performed. The
Schr\"odinger evolution operator is constructed, the trace is taken, 
and the partition function of the canonical ensemble
is obtained. The black hole entropies differ, in
general, from the usual quarter of the horizon area due to the
dilaton.

\end{abstract} 


\maketitle 

\section{Introduction} 

\subsection{Black hole thermodynamics} 
Hawking radiation \cite{hawking1} is a phenomenon that emerges when
one combines classical general relativity with quantum fields in a
black hole background.  It is thus a semiclassical phenomenon. The
radiation is thermal and emitted at a given temperature $T$ which
depends on the black hole parameters.  When this is linked to other
black hole properties one finds that black holes indeed have entropy
$S$ and allow a thermodynamical description \cite{bekenstein1}.

\subsection{Path integral approach to black hole thermodynamics} 
One can also get the thermodynamic properties of a black hole through
a path integral approach to quantum gravity.  In this approach one
uses Feynman's idea that the amplitude $<g_2,\Sigma_2,t_2|g_1,\Sigma_1,t_1>$
to go to a bra vector state $<g_2,\Sigma_2,t_2|$, with metric $g_2$ on
a spatial hypersurface $\Sigma_2$ 
at some generic prescribed time $t_2$, from one ket vector state,
$|g_1,\Sigma_1,t_1>$, with metric $g_1$ on a spatial hypersurface
$\Sigma_1$ at some generic prescribed time $t_1$, 
can be found from the sum of all possible classical vacuum
geometries, i.e., metric field configurations $g$, which take the
prescribed values $g_1$ and $g_2$ on the surfaces $\Sigma_1$ and
$\Sigma_2$, respectively, i.e., $<g_2,\Sigma_2,t_2|g_1,\Sigma_1,t_1>=\int
D[g]\,\exp\left(iS[g]\right)$, where $D[g]$ is a measure on the space
of all the gravitational field configurations, and $S$ is the
Lorentzian action of the theory. 
Now, it is known that if one
Euclideanizes the prescribed 
time $t$ such that $t=-i\beta$ and Euclideanizes the action
$I=-iS$, one obtains a timeless amplitude, which is equivalent to the
partition function $Z=\int D[g]\,\exp\left(-S[g]\right)$.  
The partition function is then computed in the saddle point
approximation, which finally connects to thermodynamics through the
relation between $Z$ and the Helmholtz free energy $F$, i.e., 
$F=-{\textbf T}\ln Z$, 
where ${\textbf T}=1/\beta$ is the temperature, and 
we put Boltzmann's constant $k_{\rm B}=1$. Since one has
computed the partition function $Z$ in this formalism one is working,
a priori, in a well defined canonical ensemble, where the temperature
${\textbf T}$ is given.  However, solitary black holes are unstable against
thermal fluctuations, indeed the heat capacity is in general negative.
This means that the canonical ensemble is not defined at all and one
should work instead in a microcanonical ensemble \cite{hawking2}.  In
the microcanonical ensemble the total energy $E$ of the system, rather
than the temperature ${\textbf T}$, is given, and the ensemble is usually well
defined.  On the other hand, calculations in the canonical ensemble
are much easier than in the microcanonical. Thus, in order to stick to
the canonical ensemble in a coherent fashion, one has to go beyond a
solitary black hole, by enclosing it in an appropriate finite box.  In
this way, the path integral formalism and its inherent connection to
the canonical ensemble, makes sense, if one carefully chooses the
boundary conditions by fixing the local temperature at the surface of
the box itself, as was done for the Schwarzschild case in
\cite{york1}, and generalized to an arbitrary static field in
\cite{zaslavskii1}.  This can also be naturally extended to the grand
canonical ensemble when one includes charge \cite{braden}.  If
further, one puts a cosmological constant into the system, i.e., one
deals with an asymptotic anti-de Sitter spacetime, one can dispense
with the box, since the cosmological constant itself yields a natural
box \cite{peca}.  This approach, called the Brown-York formalism, has
been further used a number of times in various different dimensions
and for several different theories of gravity containing black holes.
One should note that since the entropy and the other laws of black
hole thermodynamics can be derived without considering matter fields,
one finds that the black hole entropy is really an intrinsic entropy
pertaining to the black hole geometry itself.

\subsection{Hamiltonian approach to black hole thermodynamics} 

There are many other approaches to calculate the entropy and the
thermodynamics of a black hole. The route we want to follow,
motivated by the success of the 
path integral approach, is to build a Lorentzian
Hamiltonian classical theory of the gravity in question, and then
obtain a Lorentzian time evolution operator in the Schr\"odinger
picture. Afterward one performs a Wick rotation from real to imaginary
time, in order to find a well defined partition function.  In more
detail, the prescription implicit in this approach is: find the
Hamiltonian of the system, calculate then the time
evolution between a final state and an initial state, i.e., between
the bra and ket vectors of those states, $<g_2,\Sigma_2,t_2|$ and
$|g_1,\Sigma_1,t_1>$, and then Euclideanize time. Here, the amplitude
to propagate to a configuration $<g_2,\Sigma_2,t_2|$ from a
configuration $|g_1,\Sigma_1,t_1>$ , is represented by
$<g_2,\Sigma_2,t_2| \exp\left(-iH(t_2-t_1)\right)|g_1,\Sigma_1,t_1>$
in the Schr\"odinger picture.  Euclideanizing time, $t_2-t_1=-i\beta$
and summing over a complete orthonormal basis of configurations $g_n$
one obtains the partition function $Z=\sum \exp\left(-\beta
E_n\right)$, of the field $g$ at a temperature $1/\beta$, where $E_n$
is the eigenenergy of the eigenstate $g_n$.  This route is based on
the Hamiltonian methods of \cite{dirac,adm,rt,kuchar}. It was developed
by Louko and Whiting in \cite{louko1} for the specific problem of
finding black hole entropies and thermodynamic properties, and further
applied in \cite{louko2,louko3,louko4,louko5} by Louko and
collaborators, in \cite{bose} by Bose and collaborators, and in
\cite{kunstatter1,kunstatter2} by Kunstatter and collaborators.  This
approach also points to an entropy where the degrees of freedom are in
the gravitational field itself, since nowhere one mentions matter
fields.

The Louko-Whiting method \cite{louko1} relies heavily on the Hamiltonian
approach of \cite{kuchar}, which in turn is an important ramification
of the Arnowitt, Deser, and Misner approach, the ADM approach
\cite{adm} (see also \cite{dirac}), when applied to the full vacuum
Schwarzschild black hole spacetime.  This spacetime is better
described by a spherically symmetric white hole plus black hole plus
two asymptotically flat regions. These regions are well pictured in a
Kruskal, or perhaps better, in a Carter-Penrose diagram.  Indeed,
Kucha\v{r} \cite{kuchar} in studying within this formalism the
Schwarzschild black hole found the true dynamical degree of freedom of
the phase space of such a spacetime, by considering the spacelike
foliations of the full manifold. This degree of freedom is represented
by one pair of canonical variables. This pair is composed of the mass
$M$ of the solution and its conjugate momentum, which physically
represents the difference between the Killing times at right and left
spatial infinities.  In \cite{louko1}, this method was adapted by
considering a spacelike foliation to the right of the future event
horizon of the solution, enabling one to find the corresponding
reduced phase space.  {}From this one can obtain a Hamiltonian $H$ and
thus the Lorentzian time evolution operator $\exp(-iHt)$ in the
Schr\"odinger picture.

This method has been applied for various theories of gravity and in
several different dimensions. In \cite{louko1}, the first paper of the
series, a vacuum Schwarzschild black hole in four dimensions in
general relativity was placed in a box, defined as a rigid timelike
frontier. The values of the metric functions were fixed both at the
horizon and at the box.  Then, in \cite{louko2}, the same procedure
was applied to a vacuum dilatonic black hole in two dimensions. Here
there was also a rigid frontier, where the values of the fields,
dilatonic and gravitational, were fixed. Afterward, in \cite{louko3},
the method was applied to the Reissner-N\"ordstrom anti-de Sitter
black hole in four-dimensional Einstein-Maxwell theory with a negative
cosmological constant.  This time there was no rigid frontier, so the
far away asymptotic properties were defined at anti-de Sitter
infinity. Next, in \cite{louko4}, the same formalism is applied to the
case of topological black hole solutions of the equations of motion
for the four-dimensional Einstein-Maxwell-anti-de Sitter theory,
leaving however part of the Hamiltonian method implicit. Finally, it
was shown in \cite{louko5} that the entropy could also be calculated
via this Hamiltonian method for five-dimensional spherically symmetric
solutions of a one parameter family included within Lovelock gravity
theory, where the action is comprised of the Ricci scalar term and the
four-dimensional Euler density, i.e. the Gauss-Bonnet term, multiplied
by an undetermined coefficient. There were some other developments.
In \cite{bose} the Brown-York approach and the Louko-Whiting approach
are compared. The main differences lie in the choice of boundary
conditions, resulting in the fact that the Hamiltonian of Brown-York
is the internal energy, whereas the Louko-Whiting Hamiltonian is the
Helmholtz free energy. In \cite{kunstatter1,kunstatter2} the
Louko-Whiting method was applied with some modifications to generic
two-dimensional dilaton-gravity theories.  Here we want to use the
Louko-Whiting method to study the entropy and thermodynamic properties
of three-dimensional black hole solutions in gravity theories with a
dilaton, a negative cosmological constant and a free parameter
$\omega$, which can be generically comprised in a three-dimensional
Brans-Dicke theory with a cosmological constant.

\subsection{Three-dimensional dilaton-gravity black holes 
and their Hamiltonian approach to thermodynamics}

There is great interest in studying three-dimensional theories of
gravity. One of the main reasons is richness of structure with one
dimension less
\cite{btz,bhtz,horowitzwelch,hornehorowitz,lemos1,sakleberlemos,ads3_bh,zaslavskii2,carlip}.  
In
fact, it is simpler to deal with certain concepts such as temperature,
entropy, and flux of radiation when one works in three dimensions, and
it also seems easier to try a three-dimensional quantum description of
the black hole system.  Theories in three dimensions, such as general
relativity, have a rich structure. Indeed, three-dimensional general
relativity with a negative 
cosmological constant has a black hole solution, the
Ba\~nados-Teitelboim-Zanelli (BTZ) black hole \cite{btz,bhtz}.  The
BTZ black hole is also a solution of string theory with a dilaton and
other fields \cite{horowitzwelch}, and there is also a three
dimensional black string solution which can be recovered from an exact
conformal field theory \cite{hornehorowitz}.  The existence of
classical solutions depends on there being a negative cosmological
constant.  A negative constant is part of the three-dimensional black
hole spacetimes structure, where the term in the action assumes a form
given by $-2\lambda^2$.  The cosmological constant term is therefore
always negative, and thus, the black hole solutions of
three-dimensional theories are usually asymptotically anti-de Sitter.
Beyond three-dimensional general relativity and effective
three-dimensional string gravity theories there are other interesting
three-dimensional theories with a cosmological constant.  One of these
can be recovered from dimensional reduction of four-dimensional
cylindrical general relativity, yielding a three-dimensional gravity
with a dilaton with a particular coupling for the kinetic dilaton term
\cite{lemos1}.  It is then natural to set up a general
three-dimensional dilaton-gravity theory, by including an $\omega$
paraneter, which yields different couplings for the kinetic term of
the dilaton. This theory is then Brans-Dicke theory in three
dimensions \cite{sakleberlemos}.
We will study in this connection the three-dimensional black hole 
solutions of \cite{sakleberlemos}, which are parametrized by the 
Brans-Dicke parameter $\omega$, where, in principle, $\omega$ can take 
any value, i.e., $\infty>\omega>-\infty$.  Since we want to perform a 
canonical Hamiltonian analysis, using an ADM formalism supplied with 
proper boundary conditions, it is necessary to pick up from the maze 
of solutions found in \cite{sakleberlemos}, only those that fulfill 
the boundary conditions we want to impose. First, we are interested 
only in solutions with horizons.  Second, we want only solutions that 
are asymptotically anti-de Sitter at infinity.  The cases of interest 
to be studied are then black holes for which $\omega\to\pm \infty$, 
$\infty>\omega>-1$, and $-\frac32>\omega>-\infty$. As in 
\cite{sakleberlemos} we choose three typical amenable cases where an 
analytical study can be done.  These are $\omega\to\infty$ (or 
equivalently $\omega\to-\infty$), $\omega=0$, and $\omega=-3$. The 
theory for which $\omega\to\infty$ is general relativity, and the 
solution is the BTZ black hole \cite{btz}. The theory for which 
$\omega=0$ is equivalent to cylindrical four-dimensional general 
relativity and the corresponding black hole was found in 
\cite{lemos1}.  The theory for which $\omega=-3$ is just a case of 3D 
Brans-Dicke theory, with a black hole solution that can be analyzed in 
this context \cite{sakleberlemos}.  
If a quantum theory only makes sense if its classical 
form can be quantized by Hamiltonian methods, 
one should pick up only solutions which can be put 
consistently in a Hamiltonian form. 

Thus, using the prescription for the canonical variables, we foliate, 
following ADM, the three-dimensional spacetime with spacelike 
hypersurfaces of equal time, whose right end is at the anti-de Sitter 
infinity, and whose left end is at the bifurcation circle of a non 
degenerate Killing horizon. These canonical variables are then 
transformed to another set of appropriate canonical variables, 
directly related to the physical parameters of the black hole 
classical solutions. In turn, these latter variables are reduced to 
the true degrees of freedom, which can be represented by one pair of 
new canonical variables, the mass $M$ and its conjugate momentum 
$P_M$.  All this construction is limited to the right static region of 
the Carter-Penrose diagram of spacetime.  This reduced theory is then 
quantized canonically.  From the quantum theory obtained, one builds a 
time evolution operator in the Schr\"odinger picture. Upon 
continuation of this operator to imaginary time, inserting it into two 
generic base state vectors and taking the trace one obtains the 
canonical partition function, appropriate to a canonical ensemble. 
Note one needs to pay attention to the boundary conditions of the 
ensemble when trying to build an appropriate one for the geometries of 
a quantum theory of gravity. These have to respect the stability 
properties of the semiclassical approximations of this quantum theory. 
Here these suitable boundary conditions are ensured by the fact that 
the black hole solutions are asymptotically anti-de Sitter.  Now, the 
partition function obtained, given the right conditions, is dominated 
by classical Euclidean solutions.  Following \cite{louko3} we choose a 
renormalized Hawking temperature, which is taken as a fixed quantity 
in the canonical ensemble, due to the fact that the Hawking 
temperature of the black hole goes to zero at infinity.  The entropy
of the system can then be found using this whole formalism. We note
that the entropy of black holes in three-dimensional dilaton-gravity
theories of the kind we study, have been worked out in \cite{ads3_bh}
through a completely different approach, namely, using known
properties of two-dimensional conformal field theories.  As well, the
entropy and thermodynamics of the BTZ black hole have been
exhaustively studied using a number of different methods, see
\cite{zaslavskii2} for a first study and \cite{carlip} for a review.

\subsection{Structure of the paper} 
The structure of the paper is as follows. In Sec. \ref{bhsolutions} we 
present the classical solutions of the three-dimensional dilatonic 
black holes, whose quantization through Hamiltonian methods we will 
perform.  There is a free parameter $\omega$ for which we choose three 
different values, $\omega=\infty,\,0,\,-3$, corresponding to the 
BTZ black hole, the 
dimensionally reduce four-dimensional cylindrical black hole, 
and a three-dimensional dilatonic black hole, 
respectively. In Sec. \ref{whichallowadm} we 
introduce the spacetime foliation through which we will define the 
canonical coordinates and which will allow us to write the action as a 
sum of constraints multiplied by their respective Lagrange 
multipliers. Then follow three sections, Secs. \ref{btz}, \ref{zero}, 
and \ref{menostres}, where we develop the thermodynamic Hamiltonian 
formalism for the $\omega=\infty,\,0,\,-3$ black holes, respectively. 
In each section, we first give the metric and the dilaton fields 
(subsection {\bf A}).  Then we replace in the action of the theory the 
foliation ansatz, with the canonical coordinates, and determine the 
conjugate momenta of the coordinates. From there we determine the form 
of the Hamiltonian action through a Legendre transformation. It is 
then necessary to guarantee that there is a well defined variational 
principle, so as to allow the derivation of the equations of motion of 
the coordinates and respective momenta. This entails adding surface 
terms to the original action, which in turn forces the definition of 
asymptotic properties for the coordinates, momenta, and Lagrange 
multipliers. It is also necessary to fix some variables at the 
boundaries in order to have a well defined variational 
principle. After this is accomplished, we write the usual metric 
functions of a three-dimensional spherically symmetric, static black 
hole spacetime as functions of the canonical variables (subsection 
{\bf B}).  We then choose a new set of canonical variables and perform 
a canonical transformation. This new set has a physical meaning, i.e., 
one is the mass parameter of the black hole, and the other, the 
momentum conjugate to the mass, is the spatial derivative of the 
Killing time.  We rewrite the action with the new variables and obtain 
a new pair of constraints with a redefined pair of Lagrange 
multipliers (subsection {\bf C}).  The new action is then reduced to 
its true degrees of freedom, with one pair of canonical variables, the 
mass and its conjugate momentum. The reduced Hamiltonian is nothing 
more than the initial surface terms written as functions of the new 
variables (subsection {\bf D}). We quantize the theory canonically, 
defining a Hilbert space and an inner product therein, with the 
Hamiltonian now being an operator in this Hilbert space. We write the 
time evolution operator in the Schr\"odinger picture (subsection {\bf 
E}).  In order to obtain a thermodynamic description of the system we 
build a partition function in the canonical ensemble with fixed 
temperature by continuing the time evolution operator to imaginary 
time, and taking the trace in the mass eigenstates. This requires that 
we normalize the trace. Finally, through saddle point methods we 
obtain the thermodynamic functions (subsection {\bf F}). In 
Sec. \ref{conclusions} we discuss and conclude. Throughout the paper 
we make $\hbar=1$, $k_{\rm B}=1$, $c=1$, and $G=\frac{1}{8}$. 

\section{The 3D black hole solutions which allow 
a proper thermodynamic Hamiltonian description} 
\label{bhsolutionswhich} 

\subsection{The 3D black hole solutions} 
\label{bhsolutions} 

Three-dimensional black hole theories have been studied 
in, e.g., 
\cite{btz,bhtz,horowitzwelch,hornehorowitz,lemos1,sakleberlemos,ads3_bh,zaslavskii2,carlip}.
A general action that 
incorporates most of these black holes is a 
Brans-Dicke action, with  gravitational and dilaton fields 
and a cosmological constant. It is given by \cite{sakleberlemos} 
\begin{equation} 
S = \frac{1}{2\pi}\int d^3x \,\sqrt{-g}\, e^{-2\phi}\,(R-4\omega 
(\partial \phi)^2 + 4\lambda^2)+\bar{B}\,, 
\label{3dbdaction} 
\end{equation} 
where $g$ is the determinant of the three-dimensional metric 
$g_{\mu\nu}$, $R$ is the curvature scalar, $\phi$ is a scalar dilaton 
field, $\lambda$ is the cosmological constant, $\omega$ is the 
Brans-Dicke parameter, and $\bar{B}$ is a generic surface term.

The general solution for a static spherically symmetric 
metric, i.e., circularly symmetric (since we deal with tree dimensions),
is \cite{sakleberlemos} 
\begin{eqnarray} 
\label{solutions} 
ds^2 &=& -\left[ (aR)^2 - \frac{b}{(aR)^{\frac{1}{\omega+1}}} \right] 
dT^2 + \frac{dR^2}{(aR)^2 - 
\frac{b}{(aR)^{\frac{1}{\omega+1}}}} + 
R^2 d\varphi^2, \qquad \,\, \omega\neq-2\,,-\frac{3}{2}\,,-1\,, 
\label{solutions_1}\\ 
ds^2 &=& - (R^2-R)\,dT^2 + (R^2-R)^{-1} dR^2 + R^2 d\varphi^2, \qquad 
\qquad \qquad \qquad \omega=-2\,,\\ 
ds^2 &=& 4\lambda^2 R^2 \ln(bR) dT^2 - \frac{dR^2}{4\lambda^2 R^2 
\ln(bR)}  R^2 d\varphi^2, \qquad \qquad \qquad \qquad 
\quad\,\omega=-\frac{3}{2}\,,\\
ds^2 &=& -dT^2 + dR^2 + d\varphi^2, \qquad \qquad \qquad \qquad 
\qquad \qquad \qquad \qquad \qquad \,\,\omega=-1, 
\label{mink}
\end{eqnarray} 
where $T,R$ are Schwarzschild coordinates, 
$a$ is a constant related to the cosmological constant 
(see below), and $b$ is a constant of 
integration (see below), and 
the general solution for $\phi$ is given by 
\begin{eqnarray} 
\phi &=& -\frac{1}{2(\omega+1)}\ln(a\,R), \qquad \omega\neq-1\,, 
\label{phisolutions_1}\\ 
\phi &=& \textrm{constant}\,, \qquad \qquad \qquad \omega=-1\,. 
\end{eqnarray} 
For the constant $a$ one has 
\begin{eqnarray} 
a &=& \frac{2\left|(\omega+1)\lambda\right|}{\sqrt{\left|(\omega+2) 
(2\omega+3)\right|}}\,,\qquad \omega\neq-2\,,-\frac{3}{2}\,,-1\,, 
\label{a_1} \\ 
a &=& 1\,, \qquad \qquad \qquad \qquad \qquad \omega=-2,-\frac{3}{2}\,, 
\label{a_2}\\ 
a &=& 0\,, \qquad \qquad \qquad \qquad \qquad \omega=-1\,. 
\label{a_3} 
\end{eqnarray} 
The constant $b$ is related to the ADM mass of the solutions by 
\begin{eqnarray} 
M &=& \frac{\omega+2}{\omega+1}\,b\,, \qquad \,\,\, 
\omega\neq-2\,,-\frac{3}{2}\,,-1\,,\\ 
M &=& 0, \qquad \qquad \quad \,\, \omega=-2,\,-1,\\ 
M &=& -4\lambda^2 \ln{b}, \qquad \omega=-\frac{3}{2}\,. 
\end{eqnarray} 

Since we want to perform a canonical Hamiltonian analysis, 
using an ADM formalism supplied with proper boundary conditions, 
it is necessary to pick up only those solutions 
that fulfill the conditions we want to impose. First, 
we are interested only in solutions with horizons, so 
we take $b$ to be positive. 
Second, apart from a measure zero of solutions, all solutions 
have a non-zero $|\lambda|$. This does not mean  straight away 
that the solutions are asymptotically anti-de Sitter. Some have 
one type or another of singularities at infinity, which 
do not allow an imposition of proper boundary conditions. 
So, from \cite{sakleberlemos} with the corresponding Carter-Penrose 
diagrams, we discard the following solutions: 
$\omega=-1$ which is simply the Minkowski solution of a low-energy 
limit of string theory, $-1>\omega>-\frac32$ since it gives 
weird conical singularities at Carter-Penrose infinity, 
and $\omega=-\frac32$ since all the Carter-Penrose boundary 
is singular. Thus, the cases of interest to be studied are 
black holes for which $\omega\to\infty$, $\infty>\omega>-1$, and 
$-\frac32>\omega>-\infty$. For $b$ positive these solutions 
have ADM mass $M$ positive, so well defined horizons. 
As in \cite{sakleberlemos} 
we choose three typical amenable cases where 
an analytical study can be done. 
These are $\omega\to\infty$ (or equivalently $\omega\to-\infty$), 
$\omega=0$, and $\omega=-3$. The theory for which 
$\omega\to\infty$ is general relativity, and the solution 
is the BTZ black hole \cite{btz,bhtz}. The theory for which 
$\omega=0$ is equivalent to cylindrical 
four-dimensional general relativity 
and the corresponding black hole was found in \cite{lemos1}. 
The theory for which $\omega=-3$ is just a case of 3D Brans-Dicke theory, 
with a particular form for the kinetic dilaton term, 
which has a black hole solution that can be analyzed in this context 
\cite{sakleberlemos}. 

\subsection{ADM form of the metric} 
\label{whichallowadm} 

The ansatz for the metric and dilaton fields 
with which we start our canonical analysis is given by 
\begin{equation} 
ds^2 = -N(t,r)^2 dt^2+\Lambda(t,r)^2(dr+N^r(t,r) 
dt)^2+R(t,r)^2d\varphi^2\,, 
\label{ADM_ansatz} 
\end{equation} 
\begin{equation} 
{\rm e}^{-2\phi} = \left(a\,R(t,r)\right)^{\frac{1}{\omega+1}}\,. 
\label{ADMphiansatz} 
\end{equation} 
This is the ADM ansatz \cite{adm} for the metric of spherically
symmetric solutions of the three-dimensional Brans-Dicke theory. In
this we follow the basic formalism developed by Kucha\v{r}
\cite{kuchar}.  The canonical coordinates $R,\,\Lambda$ are functions
of $t$ and $r$, $R=R(t,r),\,\Lambda=\Lambda(t,r)$. Now, $r=0$ is
generically on the horizon as analyzed in \cite{kuchar}, but for our
purposes $r=0$ represents the horizon bifurcation point of the
Carter-Penrose diagram \cite{louko1} (see also
\cite{louko2}-\cite{bose}).  In three spacetime dimensions the point
represents a circle.  The coordinate $r$ tends to $\infty$ as the
coordinates themselves tend to infinity, and $t$ is another time
coordinate. The remaining functions are the lapse $N=N(t,r)$ and shift
functions $N^r=N^r(t,r)$ and will play the role of Lagrange
multipliers of the Hamiltonian of the theory.  The canonical
coordinates $R=R(t,r),\,\Lambda=\Lambda(t,r)$ and the lapse function
$N=N(t,r)$ are taken to be positive.  The angular coordinate is left
untouched, due to spherical (i.e, circular) symmetry.  The dilaton is
a simple function of the radial canonical coordinate, and it can be
traded directly by it through equation (\ref{ADMphiansatz}), as will
be done below.  The ansatz (\ref{ADM_ansatz})-(\ref{ADMphiansatz}) is
written in order to perform the foliation of spacetime into spacelike
hypersurfaces, and thus separates the spatial part of the spacetime
from the temporal part.  Indeed, the canonical analysis requires the
explicit separation of the time coordinate from the other space
coordinates, and so in all expressions time is treated separately from
the other coordinates. It breaks explicit, but not implicit,
covariance of the three-dimensional Brans-Dicke theory.  Such a split
is necessary in order to perform the Hamiltonian analysis.  The metric
coefficients of the induced metric on the hypersurfaces become the
canonical variables, and the momenta are determined in the usual way,
by replacing the time derivatives of the canonical variables, the
velocities.  Then, using the Hamiltonian one builds a time evolution
operator to construct an appropriate thermodynamic ensemble for the
geometries of a quantum theory of gravity.  Assuming that a quantum
theory only makes sense if its classical form can be quantized by
Hamiltonian methods, one should pick up only solutions which can be
put consistently in a Hamiltonian form.  Thus, in the following we
perform a Hamiltonian analysis to extract the entropy and other
thermodynamic properties in the three three-dimensional Brans-Dicke
black holes mentioned above, those for which $\omega=\infty, 0, -3$.

\section{Hamiltonian thermodynamics of the BTZ black hole 
({\large$\omega=\infty$})} 
\label{btz} 

\subsection{The metric} 
\begin{figure} 
[htmb] 
\centerline{\includegraphics 
{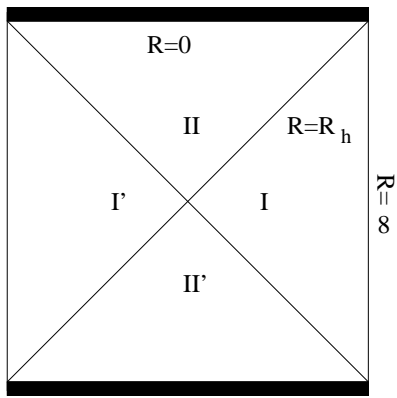}} 
\caption {{\small The Carter-Penrose Diagram for the $\omega=\infty$ 
case.}} 
\label{omega_infinity} 
\end{figure}

For $\omega\rightarrow\infty$, the three-dimensional Brans-Dicke theory 
reduces to three-dimensional general relativity \cite{sakleberlemos}. 
Then the general metric and dilaton solutions, given in 
Eqs. (\ref{solutions_1}) and (\ref{phisolutions_1}), reduce to the 
following 
\begin{eqnarray} 
ds^2 &=& - \left[ (a\,R)^2 - M \right]\,dT^2 + \frac{dR^2}{(a\,R)^2 - M} 
+R^2\,d\varphi^2\,, \label{metricbtz}\\ 
e^{-2\phi} &=& 1\,, \label{dilatonbtz} 
\end{eqnarray} 
with $M=b$ and $a=\sqrt{2}|\lambda|$. This is the BTZ black hole solution. 
Next, in Fig. \ref{omega_infinity}, we have the Carter-Penrose diagram 
of the BTZ black hole, where $R_{\rm{h}}$ denotes the black hole 
horizon radius, $R_{\rm{h}}=\sqrt{M/(2\lambda^2)}$, 
$R=0$ is the radius of the causal singularity, and $R=\infty$ 
is the spatial infinity, and we have discriminated the static and 
dynamical regions by roman numerals, namely the static right I and 
left I', and the dynamical future II and past II' regions. 

\subsection{Canonical formalism} 

The action (\ref{3dbdaction}) 
with $\omega\rightarrow\infty$, and $2\lambda^2=l^{-2}$, 
where $l$ is the AdS length, becomes, excluding surface terms, 
\begin{eqnarray} 
S[\Lambda,\,R,\,\dot{\Lambda},\,\dot{R};\,N,\,N^r] &=& \int dt 
\int_0^{\infty} dr 
\left\{- 2N^{-1}\dot{\Lambda}\dot{R} + 2 N^{-1}N^r R'\dot{\Lambda} + 
2 N^{-1}(N^r)'\Lambda \dot{R} + 2 N^{-1}N^r \Lambda ' \dot{R}\right. 
\nonumber \\ 
&& \left. - 2N^{-1}(N^r)^2\Lambda 'R' + 2N\Lambda^{-2}\Lambda 
'R'-2N\Lambda^{-1}R'' 
+ 4N\lambda^2\Lambda R \right\}\,, 
\label{infty} 
\end{eqnarray} 
where $\dot{}$ means derivative with respect to time $t$ and $'$ is 
the derivative with respect to $r$, and where all the explicit 
functional dependences are omitted.  There are two ways to arrive at 
the simple form of Eq. (\ref{infty}). One is to replace the ADM ansatz 
for the metric and the field $\phi$, given in 
(\ref{ADM_ansatz})-(\ref{ADMphiansatz}), into the action 
(\ref{3dbdaction}). One obtains an expanded action, where one has 
polynomials on the canonical coordinates and the Lagrange multipliers, 
plus their derivatives, with respect both to time $t$ and to space 
$r$. Afterward, one integrates in the angular coordinate $\varphi$ 
and obtains a two-dimensional action. Of the terms composing it some 
are total derivatives which can be discarded. This may prove itself to 
be cumbersome.  The other, simpler, way to arrive 
at the form of the action (\ref{infty}), is to dimensionally 
reduce, through the Killing angle coordinate $\varphi$, the action 
before replacing the ansatz 
(\ref{ADM_ansatz})-(\ref{ADMphiansatz}). The new action is then a 
two-dimensional integral.  In this way the total derivatives are seen 
straight away, allowing one to discard all the unnecessary surface 
terms. After the reduction and integration one can replace again the 
ansatz.  Depending on the situation we use three different letters 
containing the same information but with slightly different numerical 
values. Thus, $a$, $\lambda$, and $l$ are related by $a^2 = 2\lambda^2 
= l^{-2}$, where $l$ is the AdS length.  From the 
action (\ref{infty}) one determines the canonical momenta, conjugate 
to $\Lambda$ and $R$, respectively, 
\begin{eqnarray} 
\label{p_lambda_oo} 
P_{\Lambda} &=& -2 N^{-1} \left\{\dot{R}-R'N^r\right\}\,, \\ 
P_R &=& -2 N^{-1}\left\{ \dot{\Lambda}-(\Lambda 
  N^r)'\right\}\,. \label{p_r_oo} 
\end{eqnarray} 
By performing a Legendre transformation, we obtain 
\begin{eqnarray} \label{Hamiltonian_oo} 
\mathcal{H} &=& N\left\{-\frac12 P_R P_\Lambda -2\Lambda^{-2}\Lambda'R' 
+2\Lambda^{-1}R''-4\lambda^2\Lambda R\right\}+N^r\left\{P_R 
R'-P_{\Lambda}'\Lambda \right\} 
\nonumber \\ && \equiv NH+N^rH_r\,. 
\end{eqnarray} 
Here additional surface terms have been ignored, as for now we are 
interested in the bulk terms only.  The action in Hamiltonian form is 
then 
\begin{equation} 
\label{haction_oo} 
S[\Lambda,\,R,\,P_{\Lambda},\,P_{R};\,N,\,N^r] = \int dt 
\int_0^{\infty} dr \left\{ P_{\Lambda}\dot{\Lambda}+P_R\dot{R}-NH-N^rH_r 
\right\}\,. 
\end{equation} 
The equations of motion are 
\begin{eqnarray} 
\dot{\Lambda} &=& -\frac12 N P_R + (N^r \Lambda)'\,, \\ 
\dot{R} &=& -\frac12 N P_\Lambda - N^r R'\,, \\ 
\dot{P_R} &=& 4\lambda^2 N \Lambda - (2N'\Lambda^{-1})' + (N^r P_R)'\,,\\ 
\dot{P_{\Lambda}}&=& 4\lambda^2 N R - 2N'\Lambda^{-2}R'+N^r(P_\Lambda)'\,. 
\end{eqnarray} 
In order to have a well defined variational principle, we need to 
eliminate the surface terms of the original bulk action, which render 
the original action itself ill defined when one seeks a correct 
determination of the equations of motion through variational methods. 
These surface terms are eliminated through judicious choice of extra 
surface terms which should be added to the action.  The action 
(\ref{haction_oo}) has the following extra surface terms, after 
variation 
\begin{equation} 
\label{general_surf_terms_oo} 
\textrm{Surface terms}=\left.\left(-N^rP_R\delta R+N^r\Lambda \delta 
P_\Lambda 
+2N\Lambda^{-2}R'\delta\Lambda-2N\Lambda^{-1}\delta 
R'+2N'\Lambda^{-1}\delta R 
\right)\right|_0^\infty\,. 
\end{equation} 
In order to evaluate this expression, we need to know the asymptotic 
conditions of each 
of the above functions individually, which are functions of $(t,r)$. 

Starting with the limit $r\rightarrow 0$ we assume 
\begin{eqnarray} 
\label{falloff_0_i_oo} 
\Lambda(t,r) &=& \Lambda_0+O(r^2)\,,\\ 
R(t,r)&=& R_0+R_2r^2+O(r^4)\,,\\ 
P_\Lambda(t,r) &=& O(r^3)\,,\\ 
P_R(t,r) &=& O(r)\,,\\ 
N(t,r) &=& N_1(t)r+O(r^3)\,,\\ 
N^r(t,r) &=& O(r^3)\,. \label{falloff_0_f_oo} 
\end{eqnarray} 
With these conditions, we have for the surface terms at $r=0$ 
\begin{equation}\label{surface_0_oo} 
\left.\textrm{Surface terms}\right|_{r=0} = -2 N_1\Lambda_0^{-1}\delta 
R_0\,. 
\end{equation} 
Note that there are time dependences on the left hand side of the 
falloff conditions and that there are no such dependences on the right 
hand side, in the lower orders of the expansion in $r$. This apparent 
discrepancy stems from the fact that there is in fact no time 
dependence in the lower orders of the majority of the functions, but 
there may still exist such dependence for higher orders. Nevertheless, 
terms such as $R_0$ are functions, independent of $(t,r)$, thus 
constant, but undetermined. Their variation makes sense, as we may 
still vary between different values for these constant functions. 

For $r\rightarrow\infty$ we have 
\begin{eqnarray} \label{falloff_inf_i_oo} 
\Lambda(t,r) &=& lr^{-1}+l^3\eta(t)r^{-3} + O^\infty(r^{-5})\,,\\ 
R(t,r) &=& r + l^2\rho(t)r^{-1}+O^\infty(r^{-3})\,,\\ 
P_\Lambda(t,r) &=& O^\infty(r^{-2})\,\\ 
P_R(t,r) &=& O^\infty(r^{-4})\,,\\ 
N(t,r) &=& R(t,r)'\Lambda(t,r)^{-1}(\tilde{N}_+(t)+O^\infty(r^{-5}))\,,\\ 
N^r(t,r) &=& O^\infty(r^{-2})\,. \label{falloff_inf_f_oo} 
\end{eqnarray} 
Here, as usual, $l^{-2}=2\lambda^2$. 
These conditions imply for the surface terms in the limit 
$r\rightarrow\infty$ 
\begin{equation} 
\left.\textrm{Surface terms}\right|_{r\rightarrow\infty} = 
2\delta(M_+(t))\tilde{N}_+\,. 
\end{equation} 
where $M_+(t)=2(\eta(t)+2\rho(t))$. 
So, the surface term added to (\ref{haction_oo}) is 
\begin{equation} \label{surface_inf_oo} 
S_{\partial\Sigma}\left[\Lambda,R;N\right]=\int dt \left(2 
 R_0 N_1 \Lambda_0^{-1} - \tilde{N}_+ M_+\right)\,. 
\end{equation} 
What is left after varying 
this last surface term and adding it to the varied initial action 
(see Eq. (\ref{haction_oo})) is 
\begin{equation}\label{variationofsurfaceterm_oo} 
\int dt \left(2 R_0 \delta(N_1 \Lambda_0^{-1}) - \delta\tilde{N}_+ 
M_+\right)\,. 
\end{equation} 
We choose to fix $N_1 \Lambda_0^{-1}$ on the horizon ($r=0$) 
and $\tilde{N}_+$ 
at infinity. These choices make the surface variation 
(\ref{variationofsurfaceterm_oo}) disappear. 
The term $N_1 \Lambda_0^{-1}$ is the integrand of 
\begin{equation} 
n^a(t_1)n_a(t_2) = -\cosh \left(\int_{t_1}^{t_2}\,dt\,N_1(t) 
  \Lambda_0^{-1}(t) \right)\,, 
\end{equation} 
which is the rate of the boost suffered by the future unit normal to the 
constant $t$ hypersurfaces defined at the 
bifurcation circle, i.e., at $r\rightarrow0$, due to the evolution of the 
constant $t$ hypersurfaces. By fixing the integrand we are fixing the rate 
of the boost, which allows us to control the metric singularity when 
$r\rightarrow0$ \cite{louko1}. 

\subsection{Reconstruction, canonical transformation, and action} 

In order to reconstruct the mass and the time from the canonical data, 
which amounts to making a canonical transformation, we have to rewrite 
the  form of the solutions of Eqs. (\ref{metricbtz})-(\ref{dilatonbtz}). 
We follow Kucha\v{r} \cite{kuchar} for this 
reconstruction. 
We concentrate our analysis on the right static region of the 
Carter-Penrose 
diagram. In the static region, we define $F$ as 
\begin{equation}\label{f_1_oo} 
F(R(t,r))=(aR(t,r))^2 - b\,, 
\end{equation} 
and make the following substitutions 
\begin{equation} 
T=T(t,r)\,, \qquad \qquad R=R(t,r)\,, 
\end{equation} 
into the solution  (\ref{metricbtz}), getting 
\begin{eqnarray} 
ds^2 &=& -(F\dot{T}^2-F^{-1}\dot{R}^2)\,dt^2\,+ 
\,2(-FT'\dot{T}+F^{-1}R'\dot{R})\,dtdr\,+\,(-F(T')^2+F^{-1}\dot{R}^2)\,dr^2\,+ 
\,R^2d\varphi^2\,. 
\end{eqnarray} 
This introduces the ADM foliation directly into the solutions. 
Comparing it with the ADM metric (\ref{ADM_ansatz}), written in 
another form as 
\begin{equation} 
ds^2 = 
-(N^2-\Lambda^2(N^r)^2)\,dt^2\,+\,2\Lambda^2N^r\,dtdr\,+ 
\,\Lambda^2dr^2\,+\,R^2\,d\varphi^2\,, 
\end{equation} 
we can write a set of three equations 
\begin{eqnarray} 
\Lambda^2 &=& -F(T')^2+F^{-1}(R')^2\,,\label{adm_1_oo}\\ 
\Lambda^2N^r &=& -FT'\dot{T}+F^{-1}R'\dot{R}\,, \label{adm_2_oo}\\ 
N^2-\Lambda^2(N^r)^2 &=& F\dot{T}^2-F^{-1}\dot{R}^2\,. \label{adm_3_oo} 
\end{eqnarray} 
The first two equations, Eqs. (\ref{adm_1_oo}) and Eq. (\ref{adm_2_oo}), 
give 
\begin{equation} \label{shift_def_oo} 
N^r = \frac{-FT'\dot{T}+F^{-1}R'\dot{R}}{-F(T')^2+F^{-1}(R')^2}\,. 
\end{equation} 
This one solution, together  with Eq. (\ref{adm_1_oo}), give 
\begin{equation} \label{lapse_def_oo} 
N = \frac{R'\dot{T}-T'\dot{R}}{\sqrt{-F(T')^2+F^{-1}(R')^2}}\,. 
\end{equation} 
One can show that $N(t,r)$ is positive (see \cite{kuchar}). 
Next, putting Eqs. (\ref{shift_def_oo})-(\ref{lapse_def_oo}), 
into the definition of the conjugate momentum of 
the canonical coordinate $\Lambda$, given in 
Eq. (\ref{p_lambda_oo}), one finds 
the spatial derivative of $T(t,r)$ as a function of the canonical 
coordinates, i.e., 
\begin{equation} 
\label{t_linha_oo} 
-T' = \frac12 F^{-1}\Lambda P_\Lambda\,. 
\end{equation} 
Later we will see that $-T'=P_M$, as it 
will be conjugate to a new canonical coordinate $M$. 
Following this procedure to the end, we may then find the form of the 
new coordinate $M(t,r)$, as a function of $t$ and $r$. First, we 
need to know the form of $F$ as a function of the canonical pair 
$\Lambda\,,\,R$. For that, we replace back into Eq. (\ref{adm_1_oo}) the 
definition, in Eq. (\ref{t_linha_oo}), of $T'$, giving 
\begin{equation}\label{f_2_oo} 
F = 
\left(\frac{R'}{\Lambda}\right)^2-\left(\frac{P_\Lambda}{2}\right)^2\,. 
\end{equation} 
Equating this form of $F$ with Eq. (\ref{f_1_oo}), we obtain 
\begin{equation} \label{m_oo} 
M = 2\lambda^2 R^2 - F \,, 
\end{equation} 
where $F$ is given in Eq. (\ref{f_2_oo}) and $a^2=2\lambda^2$, 
see Eq. (\ref{a_1}). 
We thus have found the form of the new canonical coordinate, $M$. It 
is now a straightforward calculation to determine the Poisson bracket 
of this variable with $P_M=-T'$ and see that they are conjugate, thus 
making Eq. (\ref{t_linha_oo}) the conjugate momentum of $M$, i.e., 
\begin{equation}\label{p_m_oo} 
P_M = \frac12 F^{-1}\Lambda P_\Lambda\,. 
\end{equation} 

It is now necessary to find out the other new canonical variable which 
commutes with $M$ and $P_M$ and which guarantees, with its conjugate 
momentum, that the transformation from $\Lambda,\,R$, to $M$ and the 
new variable is canonical.  Immediately is it seen that $R$ commutes 
with $M$ and $P_M$. It is then a candidate. It remains to be seen 
whether $P_R$ also commutes with $M$ and $P_M$.  As with $R$, it is 
straightforward to see that $P_R$ does not commute with $M$ and $P_M$, 
as these contain powers of $R$ in their definitions, and 
$\left\{R(t,r),\,P_R(t,r^*) \right\}=\delta(r-r^*)$.  So rename the 
canonical variable $R$ as $R=\textrm{R}$.  We have then to find a new 
conjugate momentum to $\textrm{R}$ which also commutes with $M$ and 
$P_M$, making the transformation from 
$\left\{\Lambda,\,R;\,P_\Lambda,\,P_R\,\right\}\rightarrow\, 
\left\{M,\,\textrm{R};\,P_M,\,P_{\textrm{R}}\,\right\}$ a canonical 
one. 
The way to proceed is to look at the constraint $H_r$, which is called 
in this formalism the super-momentum. This is the constraint which 
generates spatial diffeomorphisms in all variables. Its form, in the 
initial canonical coordinates, is $H_r=-\Lambda\,P_\Lambda'+P_R\,R'$. 
In this formulation, $\Lambda$ is a spatial density and $R$ is a 
spatial scalar. As the new variables, $M$ and $\textrm{R}$, are 
spatial scalars, the generator of spatial diffeomorphisms is written 
as $H_r=P_M M'+P_{\textrm{R}} \textrm{R}'$, regardless of the 
particular form of the canonical coordinate transformation. It is thus 
equating these two expressions of the super-momentum $H_r$, with $M$ 
and $P_M$ written as functions of $\Lambda,\,R$ and their respective 
momenta, that gives us the equation for the new $P_{\textrm{R}}$. 
This results in 
\begin{eqnarray} 
P_{\textrm{R}} &=& P_R - 2\lambda^2 F^{-1}\Lambda P_\Lambda R + 
F^{-1}\Lambda^{-1} P_\Lambda R''- F^{-1}\Lambda^{-2}P_\Lambda \Lambda'R' 
\nonumber \\ 
&& - F^{-1}\Lambda^{-1} P_\Lambda' R'\,. \label{p_rnew_oo} 
\end{eqnarray} 
We have now all the canonical variables of the new set determined. For 
completeness and future use, we write the inverse transformation for 
$\Lambda$ and $P_\Lambda$, 
\begin{eqnarray} \label{inversetrans_1_oo} 
\Lambda &=& \left((\textrm{R}')^2F^{-1}-P_M^2F\right)^{\frac12}\,, \\ 
P_\Lambda &=& 2 F P_M 
\left((\textrm{R}')^2F^{-1}-P_M^2F\right)^{-\frac12}\,. 
\label{inversetrans_2_oo} 
\end{eqnarray} 
In summary, the canonical transformations are the following, 
\begin{eqnarray} \label{setcanonicaltrans_oo} 
R &=& \textrm{R}\,,\nonumber \\ 
M &=& 2\lambda^2 R^2 - F\,, 
\nonumber \\ 
P_{\textrm{R}} &=& P_R - 2\lambda^2 F^{-1}\Lambda P_\Lambda R + 
F^{-1}\Lambda^{-1} P_\Lambda R''-F^{-1}\Lambda^{-2}P_\Lambda \Lambda'R' 
\nonumber \\ 
&& - F^{-1}\Lambda^{-1} P_\Lambda' R'\,, \nonumber \\ 
P_M &=& \frac12 F^{-1}\Lambda P_\Lambda\,. 
\end{eqnarray} 

It remains to be seen that this set of transformations 
is in fact canonical. 
In order to prove that the set of equalities in expression 
(\ref{setcanonicaltrans_oo}) is canonical we start with the equality 
\begin{eqnarray} \label{oddidentity_oo} 
P_\Lambda \delta\Lambda + P_R \delta R - P_M \delta M - 
P_{\textrm{R}}\delta\textrm{R} &=& \left( \delta R \ln 
  \left|\frac{2 R'+\Lambda P_\Lambda}{2 R'- \Lambda 
      P_\Lambda}\right|\right)'+ \nonumber \\ 
&& +\, \delta\left(\Lambda P_\Lambda +  R' \ln 
 \left|\frac{2 R'-\Lambda P_\Lambda}{2 R'+ \Lambda 
      P_\Lambda}\right|\right)\,. 
\end{eqnarray} 
We now integrate expression (\ref{oddidentity_oo}) in $r$, in 
the interval from $r=0$ to $r=\infty$. The first term on the right 
hand side of Eq. (\ref{oddidentity_oo}) vanishes due to the falloff 
conditions (see Eqs. (\ref{falloff_0_i_oo})-(\ref{falloff_0_f_oo}) and 
Eqs. (\ref{falloff_inf_i_oo})-(\ref{falloff_inf_f_oo})). 
We then obtain the following expression 
\begin{eqnarray} \label{int_oddidentity_oo} 
\int_0^\infty\,dr\,\left(P_\Lambda \delta\Lambda + P_R \delta 
  R\right)-\int_0^\infty\,dr\,\left(P_M \delta M + 
P_{\textrm{R}}\delta\textrm{R} \right) &=& 
\delta\omega\,\left[\Lambda,\,R,\,P_\Lambda\right]\,, 
\end{eqnarray} 
where $\delta\omega\,\left[\Lambda,\,R,\,P_\Lambda\right]$ is a well 
defined functional, which is also an exact form. This equality shows 
that the difference between the Liouville form of 
$\left\{R,\,\Lambda;\,P_R,\,P_\Lambda\right\}$ and the Liouville form 
of $\left\{\textrm{R},\,M;\,P_{\textrm{R}},\,P_M\right\}$ is an exact 
form, which implies that the transformation of variables given by the 
set of equations (\ref{setcanonicaltrans_oo}) is canonical. 

Armed with the certainty of the canonicity of the new variables, we 
can write the asymptotic form of the canonical variables and of the 
metric function $F(t,r)$. These are, for $r\rightarrow 0$ 
\begin{eqnarray} \label{newfalloff_0_i_oo} 
F(t,r) &=& 4 R_2^2 \Lambda_0^{-2} r^2 + O(r^4)\,, \\ 
\textrm{R}(t,r) &=& R_0+R_2\,r^2+O(r^4)\,, \\ 
M(t,r) &=& 2\lambda^2 R_0^2 + \left(4\lambda^2R_0R_2- 
4R_2^2\Lambda_0^{-2}\right) \,r^2+O(r^4)\,, \\ 
P_\textrm{R}(t,r) &=& O(r)\,, \\ 
P_M(t,r) &=& O(r)\,.\label{newfalloff_0_f_oo} 
\end{eqnarray} 
For $r\rightarrow \infty$, we have 
\begin{eqnarray} \label{newfalloff_inf_i_oo} 
F(t,r) &=& 2\lambda^2\,r^2 - 2(\eta(t)+2\rho(t))+O^{\infty}(r^{-2})\,, \\ 
\textrm{R}(t,r) &=& r + (2\lambda^2)^{-1}\rho(t)\,r^{-1}+ 
O^{\infty}(r^{-3})\,, \\ 
M(t,r) &=& M_+(t) + O^{\infty}(r^{-2}) \,, \\ 
P_\textrm{R}(t,r) &=& O^\infty(r^{-4})\,, \\ 
P_M(t,r) &=& O^\infty(r^{-5})\,, \label{newfalloff_inf_f_oo} 
\end{eqnarray} 
where $M_+(t)=2(\eta(t)+2\rho(t))$, as seen before in the 
surface terms (see Eq. (\ref{surface_inf_oo})). 

We are now almost ready to write the action with the new canonical 
variables. It is now necessary to determine the new Lagrange 
multipliers. In order to write the new constraints with the new 
Lagrange multipliers, we can use the identity given by the space 
derivative of $M$, 
\begin{equation} 
M' = -\Lambda^{-1}\left( R' H +\frac12 P_\Lambda H_r \right)\,. 
\end{equation} 
Solving for $H$ and making use of the inverse transformations of 
$\Lambda$ and $P_\Lambda$, in Eqs. (\ref{inversetrans_1_oo}) and 
(\ref{inversetrans_2_oo}), we get 
\begin{eqnarray} 
\label{oldcnewvariables_1_oo} 
H &=& - \frac{M'F^{-1}\textrm{R}'+F P_M 
P_{\textrm{R}}}{\left(F^{-1}(\textrm{R}')^2-F 
    P_M^2\right)^{\frac12}}\,, \\ 
H_r &=& P_M M' + P_{\textrm{R}} 
\textrm{R}'\,.\label{oldcnewvariables_2_oo} 
\end{eqnarray} 
Following Kucha\v{r} \cite{kuchar}, the new set of constraints, totally 
equivalent to the old set $H(t,r)=0$ and $H_r(t,r)=0$ outside the 
horizon points, is $M'(t,r)=0$ and $P_{\textrm{R}}(t,r)=0$. By continuity, 
this also 
applies on the horizon, where $F(t,r)=0$. 
So we can say that the 
equivalence is valid everywhere. The new Hamiltonian, the total sum of 
the constraints, can now be written as 
\begin{equation} \label{newHamiltonian_oo} 
NH+N^rH_r = N^M M' + N^{\textrm{R}} P_{\textrm{R}}\,. 
\end{equation} 
In order to determine the new Lagrange multipliers, one has to write 
the left hand side of the previous equation, 
Eq. (\ref{newHamiltonian_oo}), and replace the constraints on that side 
by their expressions as functions of the new canonical coordinates, 
spelt out in Eqs. 
(\ref{oldcnewvariables_1_oo})-(\ref{oldcnewvariables_2_oo}). 
After manipulation, one gets 
\begin{eqnarray} \label{new_mult_1_oo} 
N^M &=& - \frac{N F^{-1} R'}{\left(F^{-1}(\textrm{R}')^2-F 
    P_M^2\right)^{\frac12}}+ N^r P_M \,, \\ 
N^{\textrm{R}} &=& - \frac{N F P_M}{\left(F^{-1}(\textrm{R}')^2-F 
    P_M^2\right)^{\frac12}}+ N^r R'\,. \label{new_mult_2_oo} 
\end{eqnarray} 
Using the inverse transformations 
Eqs. (\ref{inversetrans_1_oo})-(\ref{inversetrans_2_oo}), and the identity 
$R=\textrm{R}$, we can write the new multipliers as functions of the 
old variables 
\begin{eqnarray} \label{mult_trans_1_oo} 
N^M &=& - NF^{-1}R'\Lambda^{-1}+\frac12 N^r F^{-1} \Lambda P_\Lambda\,, \\ 
N^{\textrm{R}} &=& - \frac12 N P_\Lambda + N^r 
R'\,,\label{mult_trans_2_oo} 
\end{eqnarray} 
allowing us determine its asymptotic conditions from the original 
conditions given above. These transformations are non-singular for $r>0$. 
As before, for $r\rightarrow 0$, 
\begin{eqnarray} \label{mult_newfalloff_0_i_oo} 
N^M(t,r) &=& -\frac12 N_1(t) \Lambda_0 R_2^{-1} + O(r^2)\,,\\ 
N^{\textrm{R}}(t,r) &=& O(r^4)\,,\label{mult_newfalloff_0_f_oo} 
\end{eqnarray} 
and for $r\rightarrow\infty$ we have 
\begin{eqnarray} \label{mult_newfalloff_inf_i_oo} 
N^M(t,r) &=& -\tilde{N}_+(t) +  O^\infty(r^{-4})\,,\\ 
N^{\textrm{R}}(t,r) &=& O^\infty(r^{-1})\,. 
\label{mult_newfalloff_inf_f_oo} 
\end{eqnarray} 
The conditions 
(\ref{mult_newfalloff_0_i_oo})-(\ref{mult_newfalloff_inf_f_oo}) 
show that the transformations in 
Eqs. (\ref{mult_trans_1_oo})-(\ref{mult_trans_2_oo}) are satisfactory in 
the case of $r\rightarrow\infty$, but not for $r\rightarrow 0$. This 
is due to fact that in order to fix the Lagrange multipliers for 
$r\rightarrow\infty$, as we are free to do, we fix $\tilde{N}_+(t)$, 
which we already do when adding the surface term 
\begin{equation} 
- \int\, dt \, \tilde{N}_+ M_+ 
\end{equation} 
to the action, in order to obtain the equations of motion in the bulk, 
without surface terms. 
However, at $r=0$, we see that fixing the multiplier $N^M$ to values 
independent of the canonical variables is not the same as fixing $N_1 
\Lambda_0^{-1}$ to values independent of the canonical variables. We 
need to rewrite the multiplier $N^M$ for the asymptotic regime 
$r\rightarrow 0$ without affecting its behavior for 
$r\rightarrow\infty$. 
In order to proceed we have to make one assumption, which is that the 
expression given in asymptotic condition of $M(t,r)$, as $r\rightarrow 
0$, for the term of order zero, $M_0\equiv 
2\lambda^2R_0(t)^2$, defines $R_0$ as a function of $M_0$, 
and $R_0$ is the horizon radius function, $R_0\equiv 
R_{\textrm{h}}(M_0)$. Also, we assume that $M_0>0$. With these 
assumptions, we are working in the domain of the classical solutions. 
We can immediately obtain that the variation of 
$R_0$ is given in relation to the variation of $M_0$ as 
\begin{equation} \label{var_r_m_oo} 
\delta R_0 = (2\lambda)^{-2} R_0^{-1} \delta M_0\,, 
\end{equation} 
where, as defined above, $a^2=l^{-2}=2\lambda^2$. This 
expression will be used when we derive the equations of motion from 
the new action. 
We now define the new multiplier $\tilde{N}^M$ as 
\begin{equation} \label{new_n_m_oo} 
\tilde{N}^M = - N^M 
\left[(1-g)+gR_0\left(l^{-2}R_0^2\right)^{-1}\right]^{-1}\,, 
\end{equation} 
where $g(r)=1+O(r^2)$ for $r\rightarrow 0$ and $g(r)=O^\infty(r^{-5})$ 
for $r\rightarrow\infty$. This new multiplier, function of the old 
multiplier $N^M$, has as its properties for $r\rightarrow \infty$ 
\begin{equation} 
\tilde{N}^M(t,r) = \tilde{N}_+(t) + O^\infty(r^{-4})\,, 
\end{equation} 
and as its properties for $r\rightarrow0$ 
\begin{equation} 
\tilde{N}^M(t,r) = \tilde{N}_0^M(t) + O(r^{2})\,, 
\end{equation} 
where $\tilde{N}_0^M$ is given by 
\begin{equation} 
\tilde{N}_0^M = \lambda^2 N_1 R_0 R_2^{-1} \Lambda_0\,. 
\end{equation} 
When the constraint $M'=0$ holds, the last expression is 
\begin{equation} 
\tilde{N}_0^M =  N_1 \Lambda_0^{-1} \,. 
\end{equation} 
With this new constraint $\tilde{N}^M$, fixing $N_1 \Lambda_0^{-1}$ at 
$r=0$ or fixing $\tilde{N}^M$ is equivalent, there being no problems 
with $N^{\textrm{R}}$, which is left as determined in 
Eq. (\ref{new_mult_2_oo}). 

The new action is now written as the sum of $S_\Sigma$, the bulk action, 
and 
$S_{\partial\Sigma}$, the surface action, 
\begin{eqnarray} 
S\left[M, \textrm{R}, P_M, P_{\textrm{R}}; \tilde{N}^M, 
  N^{\textrm{R}}\right] &=& \int \,dt\, \int_0^\infty \, dr \, 
\left( P_M\dot{M} + P_\textrm{R} \dot{\textrm{R}} - 
N^{\textrm{R}}P_{\textrm{R}} + \tilde{N}^M 
\left[(1-g)+gR_0\left(l^{-2}R_0^2\right)^{-1}\right]\,M'\right)+ 
\nonumber \\ 
&&  \int \, dt \, \left(2 R_0 \tilde{N}_0^M - 
  \tilde{N}_+ M_+ \right)\,. \label{newaction_oo} 
\end{eqnarray} 
The new equations of motion are now 
\begin{eqnarray} \label{new_eom_1_oo} 
\dot{M} &=& 0\,, \\ 
\dot{\textrm{R}} &=& N^{\textrm{R}}\,, \\ 
\dot{P}_M &=& (N^M)'\,, \\ 
\dot{P}_{\textrm{R}} &=& 0\,, \\ 
M' &=& 0\,, \\ 
P_{\textrm{R}} &=& 0\,, \label{new_eom_6_oo} 
\end{eqnarray} 
where we understood $N^M$ to be a function of the new constraint, 
defined through Eq. (\ref{new_n_m_oo}). The resulting boundary terms of 
the variation of this new action, Eq. (\ref{newaction_oo}), are, 
first, terms proportional to $\delta M$ and $\delta \textrm{R}$ on the 
initial and final hypersurfaces, and, second, 
$ 
 \int \, dt \, \left(2 R_0 \delta\tilde{N}_0^M - 
  M_+ \delta\tilde{N}_+ \right)\,. 
$ 
Here we have used the expression in Eq. (\ref{var_r_m_oo}). 
The action in 
Eq. (\ref{newaction_oo}) yields the equations of motion, 
Eqs. (\ref{new_eom_1_oo})-(\ref{new_eom_6_oo}), provided that we fix the 
initial and final values of the new canonical variables and that we 
also fix the values of $\tilde{N}^M_0$ and of $\tilde{N}_+$. 
Thanks to 
the redefinition of the Lagrange multiplier, from $N^M$ to 
$\tilde{N}^M$, the fixation of those quantities, $\tilde{N}^M_0$ and 
$\tilde{N}_+$, has the same meaning it had before the 
canonical transformations and the redefinition of $N^M$. 
This same meaning is guaranteed through the use of our gauge freedom to 
choose the multipliers, and at the same time not fixing the boundary 
variations independently of the choice of Lagrange multipliers, which 
in turn allow us to have a well defined variational principle for the 
action. 

\subsection{Hamiltonian reduction} 

We now solve the constraints in order to reduce to the true dynamical 
degrees of freedom. The equations of motion 
(\ref{new_eom_1_oo})-(\ref{new_eom_6_oo}) allow us to write $M$ 
as an independent function of space, $r$, 
\begin{equation} \label{m_t_oo} 
M(t,r) = \textbf{m}(t)\,. 
\end{equation} 
The reduced action, with the constraints taken into account, is then 
\begin{equation} \label{red_action_oo} 
S 
\left[\textbf{m},\textbf{p}_{\textbf{m}};\tilde{N}_0^M,\tilde{N}_+\right] 
= \int 
dt\,\textbf{p}_{\textbf{m}} \dot{\bf{m}}-\textbf{h}\,, 
\end{equation} 
where 
\begin{equation} \label{new_p_m_oo} 
\textbf{p}_{\textbf{m}} = \int_0^\infty dr\,P_M\,, 
\end{equation} 
and the reduced Hamiltonian, $\textbf{h}$, is now written as 
\begin{equation} \label{red_Hamiltonian_oo} 
\textbf{h}(\textbf{m};t) = -2 R_{\textrm{h}} \tilde{N}_0^M + 
\tilde{N}_+ \textbf{m}\,, 
\end{equation} 
with $R_{\textrm{h}}$ being the horizon radius. We also have 
that $\textbf{m}>0$, according to the assumptions made in the previous 
subsection. Thanks to the functions $\tilde{N}_0^M(t)$ and 
$\tilde{N}_+(t)$ the Hamiltonian $\textbf{h}$ is an explicitly time 
dependent function. The variational principle associated with the 
reduced action, Eq. (\ref{red_action_oo}), will fix the value of 
$\textbf{m}$ on the initial and final hypersurfaces, or in the spirit 
of the classical analytical mechanics, the Hamiltonian principle fixes 
the initial and final values of the canonical coordinate. 
The equations of motion are 
\begin{eqnarray} \label{red_eom_1_oo} 
\dot{\textbf{m}} &=& 0\,, \\ 
\dot{\textbf{p}}_{\textbf{m}} &=& 2 \tilde{N}_0^M 
(4\lambda^2R_{\textrm{h}})^{-1} - \tilde{N}_+\,. 
\label{red_eom_2_oo} 
\end{eqnarray} 
The equation of motion for $\textbf{m}$, Eq. (\ref{red_eom_1_oo}), is 
understood as saying that $\textbf{m}$ is, on a classical solution, 
equal to the mass parameter $M$ of the solution, Eq. (\ref{metricbtz}). 
In order to interpret the other equation of motion, 
Eq. (\ref{red_eom_2_oo}), we have to recall that from 
Eq. (\ref{p_m_oo}) one has $P_M=-T'$, 
where $T$ is the Killing time. This, together with 
the definition of $\textbf{p}_{\textbf{m}}$, given in 
Eq. (\ref{new_p_m_oo}), yields 
\begin{equation} 
\textbf{p}_{\textbf{m}} = T_0 - T_+\,, 
\end{equation} 
where $T_0$ is the value of the Killing time at the left end of the 
hypersurface of a certain $t$, and $T_+$ is the Killing time at 
spatial infinity, the right end of the same hypersurface of $t$. As 
the hypersurface evolves in the spacetime of the black hole solution, 
the right hand side of Eq. (\ref{red_eom_2_oo}) is equal 
to $\dot{T}_0-\dot{T}_+$. 

\subsection{Quantum theory and partition function} 

The next step is to quantize the reduced Hamiltonian theory, by 
building the time evolution operator quantum mechanically and then 
obtaining a partition function through the analytic continuation of 
the same operator \cite{louko1}-\cite{louko5}. 
The variable $\textbf{m}$ is regarded here as a configuration 
variable. This variable satisfies the inequality $\textbf{m}>0$. The 
wave functions will be of the form $\psi(\textbf{m})$, with the inner 
product given by 
\begin{equation} 
\left(\psi,\chi \right) = \int_A \mu d\textbf{m}\, \bar{\psi}\chi\,, 
\end{equation} 
where $A$ is the domain of integration defined by $\textbf{m}>0$ and 
$\mu(\textbf{m})$ is a smooth and positive weight factor for the 
integration measure. It is assumed that $\mu$ is a slow varying 
function, otherwise arbitrary. We are thus working in the Hilbert 
space defined as $\mathscr{H}:=L^2(A;\mu d\textbf{m})$. 

The Hamiltonian operator, written as $\hat{\textbf{h}}(t)$, acts 
through pointwise multiplication by the function 
$\textbf{h}(\textbf{m};t)$, which on a function of our working Hilbert 
space reads 
\begin{equation} 
\hat{\textbf{h}}(t)\psi(\textbf{m})=\textbf{h}(\textbf{m};t) 
\psi(\textbf{m})\,. 
\end{equation} 
This Hamiltonian operator is an unbounded essentially self-adjoint 
operator. The corresponding time evolution operator in the same 
Hilbert space, which is unitary due to the fact that the Hamiltonian 
operator is self-adjoint, is 
\begin{equation} 
\hat{K}(t_2;t_1) = \exp 
\left[-i\int_{t_1}^{t_2}dt'\,\hat{\textbf{h}}(t') \right]\,. 
\label{K} 
\end{equation} 
This operator acts also by pointwise multiplication in the Hilbert 
space. 
We now define 
\begin{eqnarray} \label{t_oo} 
\mathcal{T} &:=& \int_{t_1}^{t_2}dt\,\tilde{N}_+(t)\,,\\ 
\Theta &:=& \int_{t_1}^{t_2}dt\,\tilde{N}^{M}_0 (t)\,. \label{theta_oo} 
\end{eqnarray} 
Using (\ref{red_Hamiltonian_oo}), (\ref{K}), (\ref{t_oo}), 
and (\ref{theta_oo}), 
we write the function $K$, which is in fact the 
action of the operator in the Hilbert space, as 
\begin{equation} 
K\left(\textbf{m};\mathcal{T},\Theta\right) = \exp 
\left[-i\textbf{m}\mathcal{T}+ 2\,i\,R_{\textrm{h}}\Theta 
\right]\,. 
\label{K2} 
\end{equation} 
This expression indicates that $\hat{K}(t_2;t_1)$ depends on $t_1$ and 
$t_2$ only through the functions $\mathcal{T}$ and $\Theta$. Thus, the 
operator corresponding to the function $K$ can now be written as 
$\hat{K}(\mathcal{T};\Theta)$. The composition law in time 
$\hat{K}(t_3;t_2)\hat{K}(t_2;t_1)=\hat{K}(t_3;t_1)$ can be regarded as 
a sum of the parameters $\mathcal T$ and $\Theta$, inside the 
operator $\hat{K}(\mathcal{T};\Theta)$. These parameters are 
evolutions parameters defined by the boundary conditions, i.e., 
$\mathcal{T}$ is the Killing time elapsed at right spatial infinity 
and $\Theta$ is the boost parameter elapsed at the bifurcation circle. 

\subsection{Thermodynamics} 

We can now build the partition function for this system. The path to 
follow is to continue the operator to imaginary time and take the 
trace over a complete orthogonal basis. 
Our classical thermodynamic situation consists of a three-dimensional 
spherically symmetric black hole, asymptotically anti-de Sitter, in 
thermal equilibrium with a bath of Hawking radiation. Ignoring 
back reaction from the radiation, the geometry is described by the 
solutions in Eq. (\ref{metricbtz}). Thus, 
we consider a thermodynamic ensemble in which the temperature, 
or more appropriately here, the inverse temperature 
$\beta$ is fixed. This characterizes a canonical ensemble, 
and the partition function 
$\mathcal{Z}(\beta)$ arises naturally in such an ensemble. 
To analytically continue the Lorentzian solution 
we put 
$\mathcal{T}=-i\beta$, and $\Theta-2\pi i$, this latter choice 
based on the regularity of the classical Euclidean solution. 
We arrive then at the following expression for the partition function 
\begin{equation} \label{partition_function_1_oo} 
\mathcal{Z}(\beta) = \textrm{Tr} \left[\hat{K}(-i\beta,-2\pi i) 
\right]\,. 
\end{equation} 
{} From Eq. (\ref{K2}) this is realized as 
\begin{equation} \label{partition_function_2_oo} 
\mathcal{Z}(\beta) = \int_0^{\infty} \mu\,d\textbf{m}\, 
\exp\left[-\beta \textbf{m}+ 4\pi 
R_{\textrm{h}}\right]\left\langle 
\textbf{m}|\textbf{m}\right\rangle\,. 
\label{Z1} 
\end{equation} 
Since $\left\langle \textbf{m}|\textbf{m}\right\rangle$ is equal to 
$\delta(0)$, one has to regularize Eq. (\ref{Z1}).  Following the 
procedure developed in the Louko-Whiting approach 
\cite{louko1}, this means regularizing and normalizing 
the operator $\hat{K}$ beforehand.  This leads to 
\begin{equation} \label{partition_function_3_oo} 
\mathcal{Z}_{\textrm{ren}}(\beta) = \mathcal{N} 
\int_{A} \mu\,d\textbf{m}\,\exp\left[-\beta \textbf{m}+4\pi 
R_{\textrm{h}}\right]\,, 
\end{equation} 
where $\mathcal{N}$ is a normalization factor 
and ${A}$ is the domain of integration. 
Provided the weight 
factor $\mu$ is slowly varying compared to the 
exponential in Eq. (\ref{partition_function_3_oo}), and using the 
fact that the horizon radius $R_{\textrm{h}}$ is a function of 
$\textbf{m}$, the integral in Eq. (\ref{partition_function_3_oo}) is 
convergent. Thermodynamically, $\textbf{m}$ is analogous to the energy 
of the system. 
Changing integration variables, from $\textbf{m}$ to $R_{\textrm{h}}$, 
where 
\begin{equation} \label{m_r_h_oo} 
\textbf{m} = 2\lambda^2 R_{\textrm{h}}^2 \,, 
\end{equation} 
the integral Eq. (\ref{partition_function_3_oo}) becomes 
\begin{equation} \label{partition_function_ren_oo} 
\mathcal{Z}_{\textrm{ren}}(\beta) = \mathcal{N} \int_{A'} 
\widetilde{\mu}\,dR_{\textrm{h}}\,\exp(-I_*)\,, 
\end{equation} 
where ${A'}$ is new the domain of integration after changing 
variables, and 
the function $I_*(R_{\textrm{h}})$,  a kind of an effective 
action (see \cite{york1}), is written as 
\begin{equation} \label{eff_action_oo} 
I_*(R_{\textrm{h}}):= 2 \beta\,\lambda^2 R_{\textrm{h}}^2-4\pi 
R_{\textrm{h}}\,. 
\end{equation} 
The new domain of integration, $A'$, is defined by the inequality 
$R_{\textrm{h}}\geq 0$. 
The new weight factor $\widetilde{\mu}$ includes the Jacobian of the 
transformation, which amounts to $\partial \textbf{m} /\partial 
R_{\textrm{h}}$. 
Since the weight factor $\widetilde{\mu}$ 
is slowly varying, we can estimate 
the integral of $\mathcal{Z}_{\textrm{ren}}(\beta)$ by 
the saddle point approximation. 
For that we have to calculate the critical points, i.e., 
we have to find the values of $R_{\textrm{h}}$ for which the first 
derivative of $I_*(R_{\textrm{h}})$ with respect to $R_{\textrm{h}}$ 
is zero. It happens for only one value of the domain 
\begin{eqnarray} \label{extrema_oo} 
R_{\textrm{h}}^+ &=&    \frac{\pi}{\beta\,\lambda^2}\,. 
\end{eqnarray} 
In order to find out the nature of the extremum we evaluate 
the second derivative at  the extremum. One finds 
\begin{eqnarray} 
\left.\frac{\partial^2 I_*}{\partial 
    R_{\textrm{h}}^2}\right|_{R_{\textrm{h}}^+} &=& 4 \beta\lambda^2\,. 
\label{min_oo} 
\end{eqnarray} 
As our domain starts at $R_{\textrm{h}}=0$, we have that the extremum 
located at $R_{\textrm{h}}=R_{\textrm{h}}^+$ is a minimum. 
Evaluating the action $I_*$ at $R_{\textrm{h}}^+$ one obtains 
\begin{equation} \label{eff_action_at_min_oo} 
I_*(R_{\textrm{h}}^+)= -\frac{2\pi^2}{\beta\lambda^2}<0\,. 
\end{equation} 
By Taylor expanding the action in the exponential of the integral, Eq. 
(\ref{partition_function_ren_oo}), 
\begin{equation} \label{taylor_action_oo} 
I_*(R_{\textrm{h}})= I_*(R_{\textrm{h}}^+) + \left.\frac{\partial 
I_*}{\partial 
    R_{\textrm{h}}}\right|_{R_{\textrm{h}}^+}  R_{\textrm{h}} + \frac 
12 \left.\frac{\partial^2 I_*}{\partial 
    R_{\textrm{h}}^2}\right|_{R_{\textrm{h}}^+} (R_{\textrm{h}})^2 + 
O((R_{\textrm{h}})^3)\,, 
\end{equation} 
we can separate the terms in such a way 
that we obtain the following expression for the renormalized partition 
function 
\begin{equation} \label{saddle_point_partition_function_oo} 
\mathcal{Z}_{\textrm{ren}}(\beta) = \exp \left[-I_*(R_{\textrm{h}}^+) 
\right] \mathcal{N} \int_{A'} \widetilde{\mu}\,dR_{\textrm{h}} 
\exp\left[2\beta\lambda^2R_{\textrm{h}}^2\right]\,. 
\end{equation} 
The Taylor expansion was up to second order, and evaluated at the 
critical point $R_{\textrm{h}}^+$, which makes the first order term of 
the expansion of $I_*(R_{\textrm{h}})$ disappear. The term which can 
be put outside the integral is the zero order term, which is the value 
of $I_*(R_{\textrm{h}})$ at the extremum $R_{\textrm{h}}^+$. The term 
left inside the exponential, $2\beta\lambda^2R_{\textrm{h}}^2$, 
is minus the second order term in the Taylor expansion, where all 
the higher orders have been ignored, as this is a good approximation, 
provided the weight factor is slowly varying. 
Finally, we may write the renormalized partition function as 
\begin{equation} 
\label{saddle_point_partition_function_P_oo} 
\mathcal{Z}_{\textrm{ren}}(\beta) = \textrm{P} \exp 
\left(\frac{2\pi^2}{\beta\lambda^2}\right)\,, 
\end{equation} 
where $\textrm{P}$ is given by 
\begin{equation} 
\label{prefactor_P_oo} 
\textrm{P} = \mathcal{N} \int_{A'} \widetilde{\mu}\,dR_{\textrm{h}} 
\exp\left(2\beta\lambda^2 R_{\textrm{h}}^2\right)\,. 
\end{equation} 
This $\textrm{P}$ is a slowly varying prefactor and this approximation 
is better as we move to higher values of 
$|I_*(R_{\textrm{h}}^+)|$. In the domain of integration the 
dominating contribution comes from the vicinity of 
$R_{\textrm{h}}=R_{\textrm{h}}^+$. 
Leaving the explicit dependence of the partition function on the 
$R_{\textrm{h}}^+$ we write the logarithm of 
$\mathcal{Z}_{\textrm{ren}}(\beta)$ as 
\begin{equation} \label{log_partition_function_oo} 
\ln(\mathcal{Z}_{\textrm{ren}}(\beta)) = \ln \textrm{P} - 
2\beta\,\lambda^2 (R_{\textrm{h}}^+)^2 + 4\pi R_{\textrm{h}}^+\,. 
\end{equation} 
By ignoring the prefactor's logarithm, which closer to 
$R_{\textrm{h}}^+$ is less relevant, we are able to determine the 
value of $\textbf{m}$ at the critical point, where we find that it 
corresponds to the value of the mass of the classical solutions of the 
black holes given in Eq. (\ref{metricbtz}). Thus, 
when the critical point dominates the partition function, we have 
that the mean energy $\left\langle E\right\rangle$ is given 
by 
\begin{equation} 
\left\langle E\right\rangle = -\frac{\partial}{\partial\beta}\ln 
\mathcal{Z}_{\textrm{ren}}(\beta) \approx 2 \lambda^2 
(R_{\textrm{h}}^+)^2 = \textbf{m}^+\,, 
\end{equation} 
where $\textbf{m}^+$ is obtained from Eq. (\ref{m_r_h_oo}) evaluated at 
$R_{\textrm{h}}^+$. 
Using Eq. (\ref{extrema_oo}) and Eq. (\ref{m_r_h_oo}), 
we obtain the temperature of the black hole $\textbf{T}\equiv 
\beta^{-1}$ 
\begin{equation} 
\textbf{T}=\left(\frac{\lambda^2\textbf{m}^+}{2\pi^2}\right)^\frac12\,, 
\label{temperature_oo} 
\end{equation} 
where $\textbf{m}^+$ is the function in (\ref{m_r_h_oo}) evaluated at 
$R_{\textrm{h}}^+$. 
Inverting Eq. (\ref{temperature_oo}), and writing the inverse temperature 
as $\beta$, we write the function $\textbf{m}^+(\beta)$ 
\begin{equation} 
\textbf{m}^+(\beta) = \frac{2\pi^2}{\beta^2\lambda^2}\,. 
\label{mass_oo} 
\end{equation} 
We see that 
$\partial\textbf{m}^+/\partial\beta<0$, which through 
$C=-\beta^2(\partial \left\langle E\right\rangle/\partial\beta)$ 
tells us that the system is thermodynamically stable. 
The entropy is given by 
\begin{equation} 
S = 
\left(1-\beta\frac{\partial}{\partial\beta}\right) 
(\ln\mathcal{Z}_{\textrm{ren}}(\beta)) 
\approx 4 \pi R_{\textrm{h}}^+\,. 
\end{equation} 
This is the entropy of the BTZ black hole 
(see \cite{btz} and also \cite{ads3_bh,zaslavskii2,carlip}).

\section{Hamiltonian thermodynamics of the general relativistic 
cylindrical dimensionally reduced black hole ({\large$\omega=0$})} 
\label{zero} 

\subsection{The metric} 
\begin{figure} 
[htmb] 
\centerline{\includegraphics 
{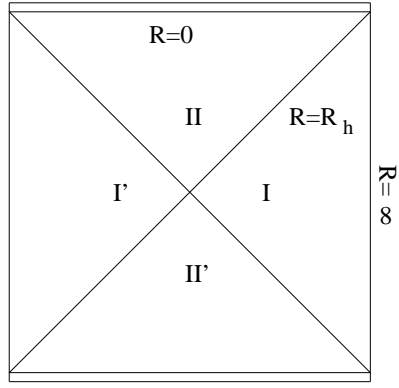}} 
\caption {{\small The Carter-Penrose Diagram for the $\omega=0$ case.}} 
\label{omega_0} 
\end{figure}

For $\omega=0$, the corresponding three-dimensional Brans-Dicke theory 
is obtained from the 
cylindrical dimensionally reduced black hole of four-dimensional 
general relativity \cite{lemos1,sakleberlemos}. 
Then general metric in Eq. (\ref{solutions}) and the 
$\phi$ field  in 
Eq. (\ref{phisolutions_1}), reduce to the following 
\begin{eqnarray} 
ds^2 &=& - \left[ (a\,R)^2 - \frac{M}{2(a\,R)} \right]\,dT^2 + 
\frac{dR^2}{(a\,R)^2 - \frac{M}{2(a\,R)}}+R^2\,d\varphi^2\,, 
\label{metric_zero}\\ 
e^{-2\phi} &=& a\,R\,,\label{phi_zero} 
\end{eqnarray} 
with $M=2b$ and $a=\sqrt{\frac23}|\lambda|$. 
Unlike the BTZ solution, 
this solution (\ref{metric_zero})-(\ref{phi_zero}) 
has a metric function whose mass term 
depends on $R$. This behavior is 
similar to the Schwarzschild black hole metric function 
\cite{lemos1}. 
In Fig. \ref{omega_0} we have plotted the Carter-Penrose diagram of the 
black 
hole solution for $\omega=0$, where again $R_{\textrm{h}}$ is the horizon 
radius, given by the larger positive real root of 
$R_{\textrm{h}}^3=\frac{1}{2\,a^3}M$, 
$R=0$ is the radius of the curvature singularity, and $R=\infty$ is the 
spatial infinity. The roman numerals mean the same as in Fig. 
\ref{omega_infinity}. 

\subsection{Canonical formalism } 
The action with $\omega=0$ becomes, excluding surface terms, 
\begin{eqnarray} 
S[\Lambda,\,R,\,\dot{\Lambda},\,\dot{R};\,N,\,N^r] &=& \int dt 
\int_0^{\infty} 
dr \,\alpha \left\{\lambda^2\Lambda N R^2 
-N^{-1}R\dot{R}\dot{\Lambda}-\frac12 N^{-1}\Lambda\dot{R}^2+N^{-1} 
R\dot{R} 
(\Lambda N^r)'\right. \nonumber\\ && 
+N^{-1}N^r R R'\dot{\Lambda} - \left. N^{-1}N^r\Lambda R R'(N^r)' 
-N^{-1}(N^r)^2RR'\Lambda'+N^{-1}N^r\Lambda\dot{R}R' \right. 
\nonumber\\ && \left. -\frac12 N^{-1}(N^r)^2\Lambda(R')^2 
-(\Lambda^{-1})'RR'N 
-\frac12 \Lambda^{-1}(R')^2N-\Lambda^{-1}RR''N\right\}\,, 
\end{eqnarray} 
where $\dot{}$ means 
derivative with respect to time $t$ and $'$ is the derivative with 
respect to $r$, and where all the explicit functional dependences are 
omitted. Depending on the situation we use four different 
letters containing the same information but with slightly different 
numerical values. Thus, $a$, $\lambda$, $l$ and $\alpha$ are 
related by $a^2=\frac23\lambda^2 = l^{-2}$, and $\alpha=4a$, 
where $l$ is, as usual, the AdS length. 
{}From this action one determines the canonical momenta, 
conjugate to $\Lambda$ and $R$ respectively 
\begin{eqnarray} \label{p_lambda} 
P_{\Lambda} &=& -\alpha N^{-1} R \left\{\dot{R}-R'N^r\right\}\,, \\ 
P_R &=& -\alpha N^{-1}\left\{ R[\dot{\Lambda}-(\Lambda 
  N^r)']+\Lambda[\dot{R}-N^rR'] \right\}\,. \label{p_r} 
\end{eqnarray} 
By performing a Legendre transformation, we obtain 
\begin{eqnarray} 
\mathcal{H} &=& N\left\{-\alpha^{-1}R^{-1}P_{\Lambda}P_{R}+\frac12 
  \alpha^{-1} \Lambda R^{-2}P_{\Lambda}^2+ 
\alpha \Lambda^{-1}RR''-\alpha\Lambda^{-2}RR'\Lambda'+\frac12 
\Lambda^{-1} (R')^2- 
\lambda^2\Lambda R^2\right\} \nonumber \\ 
&& + N^r\left\{P_R R'-P_{\Lambda}'\Lambda \right\}\equiv NH+N^rH_r\,. 
\end{eqnarray} 
The action in Hamiltonian form is then 
\begin{equation}\label{haction} 
S[\Lambda,\,R,\,P_{\Lambda},\,P_{R};\,N,\,N^r] = \int dt 
\int_0^{\infty} dr \left\{ P_{\Lambda}\dot{\Lambda}+P_R\dot{R}-NH-N^rH_r 
\right\}\,. 
\end{equation} 
The equations of motion are 
\begin{eqnarray} 
\dot{\Lambda} &=& -N\alpha^{-1}R^{-1}P_R+N\alpha^{-1}\Lambda 
R^{-2}P_{\Lambda}+N^r\Lambda \\ 
\dot{R} &=& -N\alpha^{-1}P_{\Lambda}R^{-1}+N^rR' \\ 
\dot{P_R} &=& -N\alpha^{-1}P_{\Lambda}P_RR^{-2}+N\alpha^{-1}\Lambda 
P_{\Lambda}^2R^{-3}-\left((N\alpha)'\Lambda^{-1}R\right)' 
-(N\alpha)(\Lambda^{-1}R')'+ 
2N\alpha \lambda^2\Lambda R+(N^rP_R)'\\ 
\dot{P_{\Lambda}}&=& 
-\frac12N\alpha^{-1}R^{-2}P_{\Lambda}^2-(N\alpha)'RR'\Lambda^{-2}- 
\frac12 N\alpha (R')^2\Lambda^{-2}+N\alpha\lambda^2R^2+N^rP_{\Lambda}'\,. 
\end{eqnarray} 
In order to have a well defined variational principle, we need to 
eliminate the surface terms of the original bulk action, which render 
the original action itself ill defined for a correct determination of the 
equations of motion through variational methods. 
Through the choice of added surface terms one can achieve this 
elimination. 
The action 
(\ref{haction}) 
has the following extra surface terms, after variation 
\begin{equation} 
\textrm{Surface terms}=\left.\alpha\left(-N\Lambda^{-1}R\delta 
R'+N'\Lambda^{-1}R\delta R - 
N^rP_R\delta R+N^r\Lambda \delta 
P_\Lambda+NRR'\Lambda^{-2}\delta\Lambda\right)\right|_0^\infty\,. 
\end{equation} 
In order to evaluate this expression, we need to know the asymptotic 
conditions of each of the functions of $(t,r)$. 

Starting with the limit $r\rightarrow 0$ we assume 
\begin{eqnarray} \label{falloff_0_i} 
\Lambda(t,r) &=& \Lambda_0(t)+O(r^2)\,,\\ 
R(t,r)&=& R_0(t)+R_2(t)r^2+O(r^4)\,,\\ 
P_\Lambda(t,r) &=& O(r^3)\,\\ 
P_R(t,r) &=& O(r)\,,\\ 
N(t,r) &=& N_1(t)r+O(r^3)\,\\ 
N^r(t,r) &=& N^r_1(t)r+O(r^3)\,. \label{falloff_0_f} 
\end{eqnarray} 
With these conditions, we have for the surface terms at $r=0$, 
\begin{equation}\label{surface_0} 
\left.\textrm{Surface terms}\right|_{r=0} = -\alpha 
N_1R_0\Lambda_0^{-1}\delta R_0\,. 
\end{equation} 
In the same way, for $r\rightarrow\infty$, 
\begin{eqnarray} \label{falloff_inf_i} 
\Lambda(t,r) &=& lr^{-1}+l^3\eta(t)r^{-4} + O^\infty(r^{-5})\,,\\ 
R(t,r) &=& r + l^2\rho(t)r^{-2}+O^\infty(r^{-3})\,,\\ 
P_\Lambda(t,r) &=& O^\infty(r^{-2})\,\\ 
P_R(t,r) &=& O^\infty(r^{-4})\,,\\ 
N(t,r) &=& R(t,r)'\Lambda(t,r)^{-1}(\tilde{N}_+(t)+O^\infty(r^{-5}))\,,\\ 
N^r(t,r) &=& O^\infty(r^{-2})\,. \label{falloff_inf_f} 
\end{eqnarray} 
These conditions imply for the surface terms in the limit 
$r\rightarrow\infty$, 
\begin{equation} 
\left.\textrm{Surface terms}\right|_{r\rightarrow\infty} = 
\alpha\delta(M_+)\tilde{N}_+\,, 
\end{equation} 
where $M_+(t)=\alpha(\eta(t)+3\rho(t))$. 
So, the surface term to be added to (\ref{haction}) is 
\begin{equation} \label{surface_inf} 
S_{\partial\Sigma}\left[\Lambda,R;N\right]=\int dt \left(\frac12 
\alpha R_0^2 N_1 \Lambda_0^{-1} - \tilde{N}_+ M_+\right)\,. 
\end{equation} 
What is left after varying 
this last surface term and adding it to the varied initial action 
(see Eq. (\ref{haction})) is 
\begin{equation}\label{variationofsurfaceterm} 
\int dt \left(\frac12 \alpha R_0^2 \delta(N_1 \Lambda_0^{-1}) 
 - \delta\tilde{N}_+ M_+\right)\,. 
\end{equation} 
We choose to fix $N_1 \Lambda_0^{-1}$ on 
the horizon and $\tilde{N}_+$ at infinity, which makes the surface 
variation (\ref{variationofsurfaceterm}) disappear. 

\subsection{Reconstruction, canonical transformation, and action} 

In order to reconstruct the mass and the time from the canonical data, 
which amounts to making a canonical transformation, we have to rewrite 
the general form of the solutions in Eqs. 
(\ref{metric_zero})-(\ref{phi_zero}). We again follow Kucha\v{r} 
\cite{kuchar} for the reconstruction. 
We concentrate our analysis in the right static region of the 
Carter-Penrose 
diagram. In the right static region we define $F$ as 
\begin{equation}\label{f_1} 
F(R(t,r))=(aR(t,r))^2 - \frac{b}{aR(t,r)}\,, 
\end{equation} 
and make the following substitutions 
\begin{equation} 
T=T(t,r)\,, \qquad \qquad R=R(t,r)\,, 
\end{equation} 
in the solutions above, Eqs. (\ref{metric_zero})-(\ref{phi_zero}), 
getting 
\begin{eqnarray} 
ds^2 &=& -(F\dot{T}^2-F^{-1}\dot{R}^2)\,dt^2\,+ 
\,2(-FT'\dot{T}+F^{-1}R'\dot{R})\,dtdr\,+\,(-F(T')^2+F^{-1}\dot{R}^2)\,dr^2\,+ 
\,R^2d\varphi^2\,. 
\end{eqnarray} 
This introduces the ADM foliation directly into the solutions. 
Comparing it with the ADM metric (\ref{ADM_ansatz}), written in 
another form as 
\begin{equation} 
ds^2 = 
-(N^2-\Lambda^2(N^r)^2)\,dt^2\,+\,2\Lambda^2N^r\,dtdr\,+ 
\,\Lambda^2dr^2\,+\,R^2\,d\varphi^2\,, 
\end{equation} 
we can write a set of three equations 
\begin{eqnarray} 
\Lambda^2 &=& -F(T')^2+F^{-1}(R')^2\,,\label{adm_1}\\ 
\Lambda^2N^r &=& -FT'\dot{T}+F^{-1}R'\dot{R}\,, \label{adm_2}\\ 
N^2-\Lambda^2(N^r)^2 &=& F\dot{T}^2-F^{-1}\dot{R}^2\,. \label{adm_3} 
\end{eqnarray} 
The first two equations, Eq. (\ref{adm_1}) and Eq. (\ref{adm_2}), give 
\begin{equation} \label{shift_def} 
N^r = \frac{-FT'\dot{T}+F^{-1}R'\dot{R}}{-F(T')^2+F^{-1}(R')^2}\,. 
\end{equation} 
This one solution, together  with Eq. (\ref{adm_1}), give 
\begin{equation} \label{lapse_def} 
N = \frac{R'\dot{T}-T'\dot{R}}{\sqrt{-F(T')^2+F^{-1}(R')^2}}\,. 
\end{equation} 
One can show that $N(t,r)$ is positive. 
Next, putting Eq. (\ref{shift_def}) and 
Eq. (\ref{lapse_def}) into the definition of the conjugate momentum of 
the canonical coordinate $\Lambda$, given in 
Eq. (\ref{p_lambda}), one finds the spatial derivative of $T(t,r)$ as a 
function of the canonical coordinates, i.e., 
\begin{equation}\label{t_linha} 
-T' = \alpha^{-1}R^{-1}F^{-1}\Lambda P_\Lambda\,. 
\end{equation} 
Later we will see that $-T'=P_M$, as it will be conjugate to a new 
canonical coordinate $M$. 
Following this procedure to the end, we may then find the form of the 
new coordinate $M(t,r)$, also as a function of $t$ and $r$. First, we 
need to know the form of $F$ as a function of the canonical pair 
$\Lambda\,,\,R$. For that, we replace back into Eq. (\ref{adm_1}) the 
definition of $T'$, giving 
\begin{equation}\label{f_2} 
F = \left(\frac{R'}{\Lambda}\right)^2-\left(\frac{P_\Lambda}{\alpha 
R}\right)^2\,. 
\end{equation} 
Equating this form of $F$ with Eq. (\ref{f_1}), we obtain 
\begin{equation} \label{m} 
M = \frac12 \alpha R \left(\frac{\alpha^2}{16}R^2-F\right)\,, 
\end{equation} 
where $F$ is given in Eq. (\ref{f_2}). 
We thus have found the form of the new canonical coordinate, $M$. It 
is now a straightforward calculation to determine the Poisson bracket 
of this variable with $P_M=-T'$ and see that they are conjugate, thus 
making Eq. (\ref{t_linha}) the conjugate momentum of $M$, i.e., 
\begin{equation}\label{p_m} 
P_M = \alpha^{-1}R^{-1}F^{-1}\Lambda P_\Lambda\,. 
\end{equation} 

It is now necessary to find out the other canonical variable which 
commutes with $M$ and $P_M$ and which guarantees, plus its conjugate 
momentum, that the transformation from $\Lambda,\,R$ to $M$ and the 
new variable is canonical. 
Immediately is it seen that $R$ commutes with $M$ and 
$P_M$. It is then a candidate. It remains to be seen whether $P_R$ 
also commutes with $M$ and $P_M$.  As with $R$, it is 
straightforward to see that $P_R$ does not commute with $M$ and $P_M$, 
as these contain powers of $R$ in their definitions, and 
$\left\{R(t,r),\,P_R(t,r^*) \right\}=\delta(r-r^*)$. 
So rename the new canonical variable $R$ as $R=\textrm{R}$. 
We have to find a 
new conjugate momentum to $R$ which also commutes with $M$ and $P_M$, 
making the transformation from 
$\left\{\Lambda,\,R;\,P_\Lambda,\,P_R\,\right\}\rightarrow\, 
\left\{M,\,\textrm{R};\,P_M,\,P_{\textrm{R}}\,\right\}$ a canonical 
one. The way to proceed is to look at the constraint $H_r$, 
which is called 
in this formalism the super-momentum. This is the constraint which 
generates spatial diffeomorphisms in all variables. Its form, in the 
initial canonical coordinates, is 
$H_r\,=\,P_R\,R'-\Lambda\,P_\Lambda'$.  In this formulation, $\Lambda$ 
is a spatial density and $R$ is a spatial scalar. As all new 
variables, $M$ and $\textrm{R}$, are spatial scalars, the 
generator of spatial diffeomorphisms is written as 
$H_r\,=\,P_{\textrm{R}} \textrm{R}'+\,P_M M'$, regardless of the 
particular form of the canonical coordinate transformation. It is thus 
equating these two expressions of the super-momentum $H_r$, with $M$ 
and $P_M$ written as functions of $\Lambda,\,R$ and their respective 
momenta, that gives us the equation for the new $P_{\textrm{R}}$. 
This results in 
\begin{eqnarray} 
P_{\textrm{R}} &=& P_R - \frac{3\alpha^2}{32}F^{-1}\Lambda P_\Lambda R 
- \frac12 R^{-1} \Lambda P_\Lambda + F^{-1} P_\Lambda R''\Lambda^{-1} 
- F^{-1} \Lambda^{-2} P_\Lambda \Lambda' R'+ (R')^2 F^{-1} 
\Lambda^{-1} P_\Lambda R^{-1} \nonumber \\ 
&& - F^{-1}\Lambda^{-1} P_\Lambda' R'\,.\label{p_rnew} 
\end{eqnarray} 
We have now all the canonical variables of the new set determined. For 
completeness and future use, we write the inverse transformation for 
$\Lambda$ and $P_\Lambda$, 
\begin{eqnarray} \label{inversetrans_1} 
\Lambda &=& \left((\textrm{R}')^2F^{-1}-P_M^2F\right)^{\frac12}\,, 
\\ 
P_\Lambda &=& \alpha \textrm{R} F P_M 
\left((\textrm{R}')^2F^{-1}-P_M^2F\right)^{-\frac12}\,. 
\label{inversetrans_2} 
\end{eqnarray} 
In summary, the canonical transformations are 
\begin{eqnarray} \label{setcanonicaltrans} 
R &=& \textrm{R}\,,\nonumber \\ 
M &=& \frac12 \alpha R \left(\frac{\alpha^2}{16}R^2-F\right)\,, 
\nonumber \\ 
P_{\textrm{R}} &=& P_R - \frac{3\alpha^2}{32}F^{-1}\Lambda P_\Lambda R 
- \frac12 R^{-1} \Lambda P_\Lambda + F^{-1} P_\Lambda R''\Lambda^{-1} 
- F^{-1} \Lambda^{-2} P_\Lambda \Lambda' R'+ (R')^2 F^{-1} 
\Lambda^{-1} P_\Lambda R^{-1} \nonumber \\ 
&& - F^{-1}\Lambda^{-1} P_\Lambda' R'\,, \nonumber \\ 
P_M &=& \alpha^{-1}R^{-1}F^{-1}\Lambda P_\Lambda\,. 
\end{eqnarray} 

In order to prove that the set of equalities in expression 
(\ref{setcanonicaltrans}) is canonical we start with the equality 
\begin{eqnarray} \label{oddidentity} 
P_\Lambda \delta\Lambda + P_R \delta R - P_M \delta M - 
P_{\textrm{R}}\delta\textrm{R} &=& \left( \frac12 \alpha R \delta R \ln 
  \left|\frac{\alpha R R'+\Lambda P_\Lambda}{\alpha R R'- \Lambda 
      P_\Lambda}\right|\right)'+ \nonumber \\ 
&& +\, \delta\left(\Lambda P_\Lambda + \frac12 \alpha R R' \ln 
 \left|\frac{\alpha R R'-\Lambda P_\Lambda}{\alpha R R'+ \Lambda 
      P_\Lambda}\right|\right)\,. 
\end{eqnarray} 
We now integrate expression (\ref{oddidentity}) in $r$, in 
the interval from $r=0$ to $r=\infty$. The first term on the right 
hand side of Eq. (\ref{oddidentity}) vanishes due to the falloff 
conditions (see Eqs. (\ref{falloff_0_i})-(\ref{falloff_0_f}) and 
Eqs. (\ref{falloff_inf_i})-(\ref{falloff_inf_f})). 
We then obtain the following expression 
\begin{eqnarray} \label{int_oddidentity} 
\int_0^\infty\,dr\,\left(P_\Lambda \delta\Lambda + P_R \delta 
  R\right)-\int_0^\infty\,dr\,\left(P_M \delta M + 
P_{\textrm{R}}\delta\textrm{R} \right) &=& 
\delta\omega\,\left[\Lambda,\,R,\,P_\Lambda\right]\,, 
\end{eqnarray} 
where $\delta\omega\,\left[\Lambda,\,R,\,P_\Lambda\right]$ is a well 
defined functional, which is also an exact form. This equality shows 
that the difference between the Liouville form of 
$\left\{R,\,\Lambda;\,P_R,\,P_\Lambda\right\}$ and the Liouville form 
of $\left\{\textrm{R},\,M;\,P_{\textrm{R}},\,P_M\right\}$ is an exact 
form, which implies that the transformation of variables given by the 
set of equations (\ref{setcanonicaltrans}) is canonical. 

Armed with the certainty of the canonicity of the new variables, we 
can write the asymptotic form of the canonical variables and of the 
metric function $F(t,r)$. These are, for $r\rightarrow 0$ 
\begin{eqnarray} \label{newfalloff_0_i} 
F(t,r) &=& 4 R_2(t) \Lambda_0(t)^{-2} r^2 + O(r^4)\,, \\ 
\textrm{R}(t,r) &=& R_0(t)+R_2(t)\,r^2+O(r^4)\,, \\ 
M(t,r) &=& \frac{1}{32}\alpha^2R_0(t)^3+\frac{1}{32}\alpha 
R_0(t)R_2(t)\left(3\alpha^2 
  R_0(t)-64R_2(t)\Lambda_0(t)^{-2}\right)\,r^2+O(r^4)\,, \\ 
P_\textrm{R}(t,r) &=& O(r)\,, \\ 
P_M(t,r) &=& O(r)\,.\label{newfalloff_0_f} 
\end{eqnarray} 
For $r\rightarrow \infty$, we have 
\begin{eqnarray} \label{newfalloff_inf_i} 
F(t,r) &=& \frac{\alpha^2}{16}\,r^2 - 
2(\eta(t)+2\rho(t))\,r^{-1}+O^{\infty}(r^{-2})\,, \\ 
\textrm{R}(t,r) &=& r + 16\rho(t)\alpha^{-2}r^{-2} + O^{\infty}(r^{-3})\,, 
\\ 
M(t,r) &=& M_+(t) + O^{\infty}(r^{-1}) \,, \\ 
P_\textrm{R}(t,r) &=& O^\infty(r^{-4})\,, \\ 
P_M(t,r) &=& O^\infty(r^{-6})\,, \label{newfalloff_inf_f} 
\end{eqnarray} 
where $M_+(t)=\alpha(\eta(t)+3\rho(t))$, as seen before in the 
surface terms (see Eq. (\ref{surface_inf})). 

We are now almost ready to write the action with the new canonical 
variables. It is now necessary to determine the new Lagrange 
multipliers. In order to write the new constraints with the new 
Lagrange multipliers, we can use the identity given by the space 
derivative of $M$, 
\begin{equation} 
M' = -\Lambda^{-1}\left( R' H +\alpha^{-1} R^{-1} P_\Lambda H_r \right)\,. 
\end{equation} 
Solving for $H$ and making use of the inverse transformations of 
$\Lambda$ and $P_\Lambda$, in Eqs. (\ref{inversetrans_1}) and 
(\ref{inversetrans_2}), we get 
\begin{eqnarray} \label{oldconstraintsinnewvariables_1} 
H &=& - \frac{M'F^{-1}\textrm{R}'+F P_M 
P_{\textrm{R}}}{\left(F^{-1}(\textrm{R}')^2-F 
    P_M^2\right)^{\frac12}}\,, \\ 
H_r &=& P_M M' + P_{\textrm{R}} 
\textrm{R}'\,.\label{oldconstraintsinnewvariables_2} 
\end{eqnarray} 
Following Kucha\v{r} \cite{kuchar}, the new set of constraints, totally 
equivalent to the old set $H(t,r)=0$ and $H_r(t,r)=0$ outside the 
horizon, is $M'(t,r)=0$ and $P_{\textrm{R}}(t,r)=0$. 
By continuity, this also applies on the horizon, where $F(t,r)=0$. 
So we can say that the equivalence is valid everywhere. 
So, the new Hamiltonian, the total 
sum of the constraints, can now be written as 
\begin{equation} \label{newHamiltonian} 
NH+N^rH_R = N^M M' + N^{\textrm{R}} P_{\textrm{R}}\,. 
\end{equation} 
In order to determine the new Lagrange multipliers, one has to write 
the left hand side of the previous equation, 
Eq. (\ref{newHamiltonian}), and replace the constraints on that side 
by their expressions as functions of the new canonical coordinates, 
spelt out in Eqs. 
(\ref{oldconstraintsinnewvariables_1})-(\ref{oldconstraintsinnewvariables_2}). 
After manipulation, one gets 
\begin{eqnarray} \label{new_mult_1} 
N^M &=& - \frac{N F^{-1} R'}{\left(F^{-1}(\textrm{R}')^2-F 
    P_M^2\right)^{\frac12}}+ N^r P_M \,, \\ 
N^{\textrm{R}} &=& - \frac{N F P_M}{\left(F^{-1}(\textrm{R}')^2-F 
    P_M^2\right)^{\frac12}}+ N^r R'\,. \label{new_mult_2} 
\end{eqnarray} 
Using the inverse transformations 
Eqs. (\ref{inversetrans_1})-(\ref{inversetrans_2}), and the identity 
$R=\textrm{R}$, we can write the new multipliers as functions of the 
old variables 
\begin{eqnarray} \label{mult_trans_1} 
N^M &=& - NF^{-1}R'\Lambda^{-1}+\alpha^{-1} N^r F^{-1} R^{-1} \Lambda 
P_\Lambda\,, \\ 
N^{\textrm{R}} &=& - \alpha^{-1} N R^{-1} P_\Lambda + N^r 
R'\,,\label{mult_trans_2} 
\end{eqnarray} 
allowing us to determine its asymptotic conditions from the original 
conditions given above. These transformations are non-singular for $r>0$. 
As before, for $r\rightarrow 0$, 
\begin{eqnarray} \label{mult_newfalloff_0_i} 
N^M(t,r) &=& -\frac12 N_1(t) \Lambda_0(t) R_2(t)^{-1} + O(r^2)\,,\\ 
N^{\textrm{R}}(t,r) &=& -2 N_1^r(t) R_2(t)\, r^2 + O(r^4) = 
O(r^2)\,,\label{mult_newfalloff_0_f} 
\end{eqnarray} 
and for $r\rightarrow\infty$ we have 
\begin{eqnarray} \label{mult_newfalloff_inf_i} 
N^M(t,r) &=& -\tilde{N}_+(t) +  O^\infty(r^{-4})\,,\\ 
N^{\textrm{R}}(t,r) &=& O^\infty(r^{-2})\,. 
\label{mult_newfalloff_inf_f} 
\end{eqnarray} 
These conditions 
(\ref{mult_newfalloff_0_i})-(\ref{mult_newfalloff_inf_f}) 
show that the transformations in 
Eqs. (\ref{mult_trans_1})-(\ref{mult_trans_2}) are satisfactory in 
the case of $r\rightarrow\infty$, but not for $r\rightarrow 0$. This 
is due to fact that in order to fix the Lagrange multipliers for 
$r\rightarrow\infty$, as we are free to do, we fix $\tilde{N}_+(t)$, 
which we already do when adding the surface term 
\begin{equation} 
- \int\, dt \, \tilde{N}_+ M_+ 
\end{equation} 
to the action, in order to obtain the equations of motion in the bulk, 
without surface terms. 
However, at $r=0$, we see that fixing the multiplier $N^M$ to values 
independent of the canonical variables is not the same as fixing $N_1 
\Lambda_0^{-1}$ to values independent of the canonical variables. We 
need to rewrite the multiplier $N^M$ for the asymptotic regime 
$r\rightarrow 0$ without affecting its behavior for 
$r\rightarrow\infty$. 
In order to proceed we have to make one assumption, which is that the 
expression given in asymptotic condition of $M(t,r)$, as $r\rightarrow 
0$, for the term of order zero, $M_0\equiv 
\frac{1}{32}\alpha^3R_0(t)^3$, defines $R_0$ as a function of $M_0$, 
and $R_0$ is the horizon radius function, $R_0\equiv 
R_{\textrm{h}}(M_0)$. Also, we assume that $M_0>0$. With these 
assumptions, we are working in the domain of the classical solutions. 
We can immediately obtain that the variation of 
$R_0$ is given in relation to the variation of $M_0$ as 
\begin{equation} \label{var_r_m} 
\delta R_0 = \frac{32}{3}\alpha^{-3} R_0^{-2} \delta M_0\,. 
\end{equation} 
This 
expression will be used when we derive the equations of motion from 
the new action. 
We now define the new multiplier $\tilde{N}^M$ as 
\begin{equation} \label{new_n_m} 
\tilde{N}^M = - N^M 
\left[(1-g)+ 2g (\alpha R_0) \left(\frac{3\alpha^3}{16} 
R_0^2\right)^{-1}\right]^{-1}\,, 
\end{equation} 
where $g(r)=1+O(r^2)$ for $r\rightarrow 0$ and $g(r)=O^\infty(r^{-5})$ 
for $r\rightarrow\infty$. This new multiplier, function of the old 
multiplier $N^M$, has as its properties for $r\rightarrow \infty$ 
\begin{equation} 
\tilde{N}^M(t,r) = \tilde{N}_+(t) + O^\infty(r^{-5})\,, 
\end{equation} 
and as its properties for $r\rightarrow0$ 
\begin{equation} 
\tilde{N}^M(t,r) = \tilde{N}_0^M(t) + O(r^{2})\,, 
\end{equation} 
where $\tilde{N}_0^M$ is given by 
\begin{equation} 
\tilde{N}_0^M = \frac{3}{64}\alpha^2 N_1 R_0 R_2^{-1} \Lambda_0\,. 
\end{equation} 
When the constraint $M'=0$ holds, the last expression is 
\begin{equation} 
\tilde{N}_0^M =  N_1 \Lambda_0^{-1} \,. 
\end{equation} 
With this new constraint $\tilde{N}^M$, fixing $N_1 \Lambda_0^{-1}$ at 
$r=0$ or fixing $\tilde{N}^M$ is equivalent, there being no problems 
with $N^{\textrm{R}}$, which is left as determined in 
Eq. (\ref{new_mult_2}). 

The new action is now written as the sum of $S_\Sigma$, the bulk action, 
and 
$S_{\partial\Sigma}$, the surface action, 
\begin{eqnarray} 
S\left[M, \textrm{R}, P_M, P_{\textrm{R}}; \tilde{N}^M, 
  N^{\textrm{R}}\right] &=& \int dt\,\int_0^\infty dr \, 
\left( P_M\dot{M} + P_\textrm{R} \dot{\textrm{R}} - 
N^{\textrm{R}}P_{\textrm{R}} + \tilde{N}^M 
\left[(1-g)+2gR_0
\left(\frac{3\alpha^2}{16}R_0^2\right)^{-1}\right]\,M'\right) 
\nonumber \\ 
&+&\int dt \, \left(\frac12 \alpha R_0^2 \tilde{N}_0^M - 
  \tilde{N}_+ M_+ \right)\,. \label{newaction} 
\end{eqnarray} 
The new equations of motion are now 
\begin{eqnarray} \label{new_eom_1} 
\dot{M} &=& 0\,, \\ 
\dot{\textrm{R}} &=& N^{\textrm{R}}\,, \\ 
\dot{P}_M &=& (N^M)'\,, \\ 
\dot{P}_{\textrm{R}} &=& 0\,, \\ 
M' &=& 0\,, \\ 
P_{\textrm{R}} &=& 0\,. \label{new_eom_6} 
\end{eqnarray} 
Here we understood $N^M$ to be a function of the new constraint, 
defined through Eq. (\ref{new_n_m}). The resulting boundary terms of 
the variation of this new action, Eq. (\ref{newaction}), are, first, 
terms proportional to $\delta M$ and $\delta \textrm{R}$ on the initial 
and 
final hypersurfaces, and, second, the term 
$ \int \, dt \, \left(\frac12 \alpha R_0^2 \delta\tilde{N}_0^M - 
  M_+ \delta\tilde{N}_+ \right)$. 
Here we have used the expression in Eq. (\ref{var_r_m}). The action in 
Eq. (\ref{newaction}) yields the equations of motion, 
Eqs. (\ref{new_eom_1})-(\ref{new_eom_6}), provided that we fix the 
initial and final values of the new canonical variables and that we 
also fix the values of $\tilde{N}^M_0$ and of $\tilde{N}_+$. Thanks to 
the redefinition of the Lagrange multiplier, from $N^M$ to 
$\tilde{N}^M$, the fixation of those quantities, $\tilde{N}^M_0$ and 
$\tilde{N}_+$, has the same meaning it had before the 
canonical transformations and the redefinition of $N^M$. 
This same meaning is guaranteed through the use of our gauge freedom to 
choose the multipliers, and at the same time not fixing the boundary 
variations independently of the choice of Lagrange multipliers, which 
in turn allow us to have a well defined variational principle for the 
action. 

\subsection{Hamiltonian reduction} 

We now solve the constraints in order to reduce to the true dynamical 
degrees of freedom. The equations of motion 
(\ref{new_eom_1})-(\ref{new_eom_6}) 
allow us to write $M$ as an independent function of the radial 
coordinate $r$, 
\begin{equation} \label{m_t} 
M(t,r) = \textbf{m}(t)\,. 
\end{equation} 
The reduced action, with the constraints and  Eq. (\ref{m_t}) 
taken into account, is 
\begin{equation} \label{red_action} 
S 
\left[\textbf{m},\textbf{p}_{\textbf{m}};\tilde{N}_0^M,\tilde{N}_+\right] 
= \int 
dt\,\textbf{p}_{\textbf{m}} \dot{\bf{m}}-\textbf{h}\,, 
\end{equation} 
where 
\begin{equation} \label{new_p_m} 
\textbf{p}_{\textbf{m}} = \int_0^\infty dr\,P_M\,, 
\end{equation} 
and the reduced Hamiltonian, $\textbf{h}$, is now written as 
\begin{equation} \label{red_Hamiltonian} 
\textbf{h}(\textbf{m};t) = -\frac12 \alpha R_{\textrm{h}}^2 \tilde{N}_0^M 
+ \tilde{N}_+ \textbf{m}\,, 
\end{equation} 
with $R_{\textrm{h}}$ being the horizon radius. We also have 
that $\textbf{m}>0$, according to the assumptions made in the previous 
subsection. Thanks to the functions $\tilde{N}_0^M(t)$ and 
$\tilde{N}_+(t)$ the Hamiltonian $\textbf{h}$ is an explicitly time 
dependent function. The variational principle associated with the 
reduced action, Eq. (\ref{red_action}), will fix the value of 
$\textbf{m}$ on the initial and final hypersurfaces, or in the spirit 
of the classical analytical mechanics, the Hamiltonian principle fixes 
the initial and final values of the canonical coordinate. 
The equations of motion are 
\begin{eqnarray} \label{red_eom_1} 
\dot{\textbf{m}} &=& 0\,, \\ 
\dot{\textbf{p}}_{\textbf{m}} &=& \frac{32}{3}\,\alpha^{-2} 
R_{\textrm{h}}^{-1}\tilde{N}_0^M - \tilde{N}_+\,. \label{red_eom_2} 
\end{eqnarray} 
The equation of motion for $\textbf{m}$, Eq. (\ref{red_eom_1}), is 
understood as saying that $\textbf{m}$ is, on a classical solution, 
equal to the mass parameter $M$ of the solutions in Eq. 
(\ref{metric_zero}). 
In order to interpret the other equation of motion, 
Eq. (\ref{red_eom_1}), we have to recall that from Eq. (\ref{p_m}) 
one has $P_M=-T'$, where $T$ is the Killing time. This, together with 
the definition of $\textbf{p}_{\textbf{m}}$, given in 
Eq. (\ref{new_p_m}), yields 
\begin{equation} 
\textbf{p}_{\textbf{m}} = T_0 - T_+\,, 
\end{equation} 
where $T_0$ is the value of the Killing time at the left end of the 
hypersurface of a certain $t$, and $T_+$ is the Killing time at 
spatial infinity, the right end of the same hypersurface of $t$. As 
the hypersurface evolves in the spacetime of the black hole solutions, 
the right hand side of Eq. (\ref{red_eom_2}) is equal to 
$\dot{T}_0 - \dot{T}_+$. 

\subsection{Quantum theory and partition function} 

The next step is to quantize the reduced Hamiltonian theory, by 
building the time evolution operator quantum mechanically and then 
obtaining a partition function through the analytic continuation of 
the same operator \cite{louko1}-\cite{louko5}. 
The variable $\textbf{m}$ is regarded here as a configuration 
variable. This variable satisfies the inequality $\textbf{m}>0$. The 
wave functions will be of the form $\psi(\textbf{m})$, with the inner 
product 
\begin{equation} 
\left(\psi,\chi \right) = \int_A \mu d\textbf{m}\, \bar{\psi}\chi\,, 
\end{equation} 
where $A$ is the domain of integration defined by $\textbf{m}>0$ and 
$\mu(\textbf{m})$ is a smooth and positive weight factor for the 
integration measure. It is assumed that $\mu$ is a slow varying 
function, otherwise arbitrary. We are thus working in the Hilbert 
space defined as $\mathscr{H}:=L^2(A;\mu d\textbf{m})$. 

Again, the Hamiltonian operator, written as $\hat{\textbf{h}}(t)$, acts 
through pointwise multiplication by the function 
$\textbf{h}(\textbf{m};t)$, which on a function of our working Hilbert 
space reads 
\begin{equation} 
\hat{\textbf{h}}(t)\psi(\textbf{m})=\textbf{h}(\textbf{m};t) 
\psi(\textbf{m})\,. 
\end{equation} 
This Hamiltonian operator is an unbounded essentially self-adjoint 
operator. The corresponding time evolution operator in the same 
Hilbert space, which is unitary due to the fact that the Hamiltonian 
operator is self-adjoint, is 
\begin{equation} 
\hat{K}(t_2;t_1) = \exp 
\left[-i\int_{t_1}^{t_2}dt'\,\hat{\textbf{h}}(t') \right]\,. 
\label{kappa_op_zero} 
\end{equation} 
This operator acts also by pointwise multiplication in the Hilbert 
space. 
We define again 
\begin{eqnarray} \label{t} 
\mathcal{T} &:=& \int_{t_1}^{t_2}dt\,\tilde{N}_+(t)\,,\label{t_zero}\\ 
\Theta &:=& \int_{t_1}^{t_2}dt\,\tilde{N}^{M}_0 (t)\,. \label{theta_zero} 
\end{eqnarray} 
Using (\ref{red_Hamiltonian}), (\ref{kappa_op_zero}), (\ref{t_zero}), 
and (\ref{theta_zero}), we write the function which is in fact the 
action of the operator in the Hilbert space 
\begin{equation} 
K\left(\textbf{m};\mathcal{T},\Theta\right) = \exp 
\left[-i\textbf{m}\mathcal{T}+\frac{i}{2}\alpha R^2_{\textrm{h}} \Theta 
\right]\,. 
\label{kappa_function_zero} 
\end{equation} 
This expression indicates that $\hat{K}(t_2;t_1)$ depends on $t_1$ and 
$t_2$ only through the functions $\mathcal{T}$ and $\Theta$. We can now 
write the operator corresponding to the function $K$ as 
$\hat{K}(\mathcal{T};\Theta)$. The composition law in time 
$\hat{K}(t_3;t_2)\hat{K}(t_2;t_1)=\hat{K}(t_3;t_1)$ can now be 
regarded as a sum of the parameters inside 
$\hat{K}(\mathcal{T};\Theta)$. 
These parameters are evolution parameters defined 
by the boundary conditions, i.e., $\mathcal{T}$ is the 
Killing time elapsed at right spatial infinity and $\Theta$ 
is the boost parameter elapsed at the bifurcation circle. 

\subsection{Thermodynamics} 

We can now build the partition function for this system. The path to 
follow is to continue the operator to imaginary time and take the 
trace over a complete orthogonal base.  Our classical thermodynamic 
situation consists of a three-dimensional circularly symmetric black 
hole with a particular dilaton function, asymptotically anti-de 
Sitter, in thermal equilibrium with a bath of Hawking 
radiation. Ignoring back reaction from the radiation, the geometry is 
described by the solution in 
Eqs. (\ref{metric_zero})-(\ref{phi_zero}).  Thus, we consider a 
thermodynamic ensemble in which the temperature, or more appropriately 
here, the inverse temperature $\beta$ is fixed.  This characterizes a 
canonical ensemble, and the partition function $\mathcal{Z}(\beta)$ 
arises naturally in such an ensemble.  To analytically continue the 
Lorentzian solution to imaginary time, we put $\mathcal{T}=-i\beta$ 
and $\Theta=-2\pi i$, based on the regularity of the classical 
Euclidean solution.  We arrive then at the following expression for 
the partition function 
\begin{equation} \label{partition_function_1} 
\mathcal{Z}(\beta) = \textrm{Tr} \left[\hat{K}(-i\beta,-2\pi i) 
\right]\,. 
\end{equation} 
{}From Eq. (\ref{kappa_function_zero}) this is realized as 
\begin{equation} \label{partition_function_2} 
\mathcal{Z}(\beta) = \int_0^{\infty} \mu\,d\textbf{m}\, 
\exp\left[-\beta \textbf{m}+\pi 
\alpha R_{\textrm{h}}^2\right]\left\langle \textbf{m}|\textbf{m} 
\right\rangle\,. 
\end{equation} 
Since $\left\langle \textbf{m}|\textbf{m}\right\rangle$ is equal to 
$\delta (0)$, one has to regularize (\ref{partition_function_2}). 
Again, following the Louko-Whiting procedure \cite{louko1}-\cite{louko5}, 
we have to regularize and normalize the operator 
$\hat{K}$ beforehand. This leads to 
\begin{equation} \label{partition_function_3} 
\mathcal{Z}_{\textrm{ren}}(\beta) = \mathcal{N} \int_{A} 
\mu\,d\textbf{m}\,\exp\left[-\beta \textbf{m}+\pi 
\alpha R_{\textrm{h}}^2\right]\,, 
\end{equation} 
where $\mathcal{N}$ is a normalization factor and $A$ is the domain 
of integration.  Provided the weight 
factor $\mu$ is slowly varying compared to the 
exponential in Eq. (\ref{partition_function_3}), and using the 
fact that the horizon radius $R_{\textrm{h}}$ is a function of 
$\textbf{m}$, the integral in Eq. (\ref{partition_function_3}) is 
convergent. Thermodynamically, $\textbf{m}$ is analogous to the energy 
of the system. 
Changing integration variables, from $\textbf{m}$ to $R_{\textrm{h}}$, 
where 
\begin{equation} \label{m_r_h} 
\textbf{m} = 2^{-5}\alpha^3 R_{\textrm{h}}^3 \,, 
\end{equation} 
the integral Eq. (\ref{partition_function_3}) becomes 
\begin{equation} \label{partition_function_ren} 
\mathcal{Z}_{\textrm{ren}}(\beta) = \mathcal{N} \int_{A'} 
\widetilde{\mu}\,dR_{\textrm{h}}\,\exp(-I_*)\,, 
\end{equation} 
where $A'$ is the new domain of integration after changing variables, 
and the function $I_*(R_{\textrm{h}})$, a kind of an effective action 
(see \cite{york1}), is written as 
\begin{equation} \label{eff_action} 
I_*(R_{\textrm{h}}):= 2^{-5}\beta\,\alpha^3 R_{\textrm{h}}^3-\pi\alpha 
R_{\textrm{h}}^2\,. 
\end{equation} 
The domain of integration, $A'$, is defined by the inequality 
$R_{\textrm{h}}\geq 0$. 
The new weight factor $\widetilde{\mu}$ includes the Jacobian of the 
change of variables, which amounts to $\partial \textbf{m} /\partial 
R_{\textrm{h}}$. Since the weight factor is slowly varying, 
we can estimate the integral of $\mathcal{Z}_{\textrm{ren}}(\beta)$ by 
the saddle point approximation. 
For that we have to calculate the critical points, which in our case 
amounts to finding the values of $R_{\textrm{h}}$ for which the first 
derivative of $I_*(R_{\textrm{h}})$ with respect to $R_{\textrm{h}}$ 
is null. It happens for two different values of the domain 
\begin{eqnarray} \label{extrema_0} 
R_{\textrm{h}}^- &=& 0\,,\\ 
R_{\textrm{h}}^+ &=& 2^6 \pi 
\left(3\beta\,\alpha^2\right)^{-1}\,.\label{extrema} 
\end{eqnarray} 
In order to find the local extrema we evaluate the second derivative 
at these two points. One finds 
\begin{eqnarray} \label{max} 
\left.\frac{\partial^2 I_*}{\partial 
    R_{\textrm{h}}^2}\right|_{R_{\textrm{h}}^-} &=& -2\pi\alpha\,, \\ 
\left.\frac{\partial^2 I_*}{\partial 
    R_{\textrm{h}}^2}\right|_{R_{\textrm{h}}^+} &=& 2\pi\alpha\,. 
\label{min} 
\end{eqnarray} 
Our domain starts at $R_{\textrm{h}}=0$, which is a local 
maximum. The global extremum, which is a minimum, is 
located at $R_{\textrm{h}}=R_{\textrm{h}}^+$. Evaluating the action 
$I_*$ at $R_{\textrm{h}}^+$ one obtains 
\begin{equation} \label{eff_action_at_min} 
I_*(R_{\textrm{h}}^+)= 2^{-5}\beta\,\alpha^3 
(R_{\textrm{h}}^+)^3-\pi\alpha 
(R_{\textrm{h}}^+)^2\,.\label{action_at_extrema} 
\end{equation} 
Substituting Eq. (\ref{extrema}) into Eq. (\ref{action_at_extrema}) gives 
\begin{equation} 
I_*(R_{\textrm{h}}^+)= -2^{12}3^{-3}\pi^3\beta^{-2}\alpha^{-3}\,. 
\label{I_Rh+} 
\end{equation} 
{}From Eq. (\ref{I_Rh+}) one sees that $I_*(R_{\textrm{h}}^+)<0$. 
By Taylor expanding the action in the exponential of the integral, Eq. 
(\ref{partition_function_ren}), i.e., 
\begin{equation} \label{taylor_action} 
I_*(R_{\textrm{h}})= I_*(R_{\textrm{h}}^+) + \left.\frac{\partial 
I_*}{\partial 
    R_{\textrm{h}}}\right|_{R_{\textrm{h}}^+}  R_{\textrm{h}} + \frac 
12 \left.\frac{\partial^2 I_*}{\partial 
    R_{\textrm{h}}^2}\right|_{R_{\textrm{h}}^+} (R_{\textrm{h}})^2 + 
O((R_{\textrm{h}})^3)\,, 
\end{equation} 
we can separate the terms in such a way 
that we obtain the following expression for the renormalized partition 
function 
\begin{equation} \label{saddle_point_partition_function} 
\mathcal{Z}_{\textrm{ren}}(\beta) = \exp \left[-I_*(R_{\textrm{h}}^+) 
\right] \mathcal{N} \int_{A'} \widetilde{\mu}\,dR_{\textrm{h}} 
\exp\left[-\pi\alpha R_{\textrm{h}}^2\right]\,. 
\end{equation} 
The Taylor expansion is up to second order, and evaluated at the 
critical point $R_{\textrm{h}}^+$, which makes the first order term of 
the expansion of $I_*(R_{\textrm{h}})$ disappear. The term which can 
be put outside the integral is the zero order term, which is the value 
of $I_*(R_{\textrm{h}})$ at the extremum $R_{\textrm{h}}^+$. The term 
left inside the exponential, $-\pi\alpha R_{\textrm{h}}^2$, is minus the 
second order term in the Taylor expansion, where all the higher orders 
have been ignored, as this is a good approximation, provided the 
weight factor is slowly varying. 
Finally, we may write the renormalized partition function as 
\begin{equation} \label{saddle_point_partition_function_P} 
\mathcal{Z}_{\textrm{ren}}(\beta) = \textrm{P} \exp 
[2^{12}3^{-3}\pi^3\beta^{-2}\alpha^{-3}]\,, 
\end{equation} 
where $\textrm{P}$ is given by 
\begin{equation} \label{prefactor_P} 
\textrm{P} = \mathcal{N} \int_{A'} \widetilde{\mu}\,dR_{\textrm{h}} 
\exp\left[-\pi\alpha R_{\textrm{h}}^2\right]\,. 
\end{equation} 
This $\textrm{P}$ is a slowly varying prefactor and this approximation 
is better as we move to higher values of 
$|I_*(R_{\textrm{h}}^+)|$. In the domain of integration the 
dominating contribution comes from the vicinity of 
$R_{\textrm{h}}=R_{\textrm{h}}^+$. 
Leaving the explicit dependence of the partition function on the 
$R_{\textrm{h}}^+$ we write the logarithm of 
$\mathcal{Z}_{\textrm{ren}}(\beta)$ as 
\begin{equation} \label{log_partition_function} 
\ln(\mathcal{Z}_{\textrm{ren}}(\beta)) = \ln \textrm{P} + \pi\,\alpha 
(R_{\textrm{h}}^+)^2 - 2^{-5} \beta \, \alpha^3 (R_{\textrm{h}}^+)^3\,. 
\end{equation} 
By ignoring the prefactor's logarithm, which closer to 
$R_{\textrm{h}}^+$ is less relevant, we are able to determine the 
value of $\textbf{m}$ at the critical point, where we find that it 
corresponds to the value of the mass of the classical solution of the 
black hole given in Eq. (\ref{metric_zero}). Thus, 
when the critical point dominates the partition function, we have that 
the mean energy $\left\langle E \right\rangle$ is given by 
\begin{equation} 
\left\langle E\right\rangle = -\frac{\partial}{\partial\beta}\ln 
\mathcal{Z}_{\textrm{ren}}(\beta) \approx 2^{-5} \alpha^3 
(R_{\textrm{h}}^+)^3 = \textbf{m}^+\,, 
\end{equation} 
where $\textbf{m}^+$ is obtained from Eq. (\ref{m_r_h}) evaluated at 
$R_{\textrm{h}}^+$. 
By replacing the value of $R_{\textrm{h}}^+$ in Eq. (\ref{extrema}) 
into Eq. (\ref{m_r_h}), we write the temperature 
$\textbf{T}\equiv\beta^{-1}$ 
\begin{equation} 
\textbf{T} = \left(\frac{3^{\frac32}
\left|\lambda\right|^3\textbf{m}^+}{2^{\frac52}\pi^3} 
\right)^{\frac13}\,, 
\label{temperature_0} 
\end{equation} 
where $\textbf{m}^+$ is the function in (\ref{m_r_h}) evaluated at 
$R_{\textrm{h}}^+$. 
Inverting Eq. (\ref{temperature_0}) we obtain the function 
$\textbf{m}^+(\beta)$, where $\beta^{-1}$ is the inverse temperature, 
\begin{equation} 
\textbf{m}^+(\beta) = 2^{13} \alpha^{-3} \pi^3 (3 \beta)^{-3}\,. 
\label{mass_0} 
\end{equation} 
We see that 
$\partial\textbf{m}^+/\partial\beta<0$, which through the heat capacity 
$C=-\beta^2(\partial \left\langle E\right\rangle/\partial\beta)$ 
tells us that the system is thermodynamically stable. 
The entropy is given by 
\begin{equation} 
S = 
\left(1-\beta\frac{\partial}{\partial\beta}\right) 
(\ln\mathcal{Z}_{\textrm{ren}}(\beta)) 
\approx  \pi \alpha {R_{\textrm{h}}^+}^2 = \,4\pi a 
{R_{\textrm{h}}^+}^2\,. 
\label{entropyomega=0} 
\end{equation} 
This is the entropy of the three-dimensional black hole, which is a 
solution of the three-dimensional dilaton-gravity theory obtained 
through dimensionally reduced cylindrically general relativity (see 
also \cite{ads3_bh}).  Note that the entropy obtained in 
Eq. (\ref{entropyomega=0}) is not proportional to the area (i.e, 
circumference in the three-dimensional case) as the entropy of the BTZ 
black hole discussed in Section \ref{btz}. This is certainly due to 
the presence of the coupling of the dilaton to the metric.

\section{Hamiltonian thermodynamics of a representative 
dilatonic black hole ({\large $\omega=-3$})} 
\label{menostres} 
\subsection{The metric} 
\begin{figure} 
[htmb] 
\centerline{\includegraphics 
{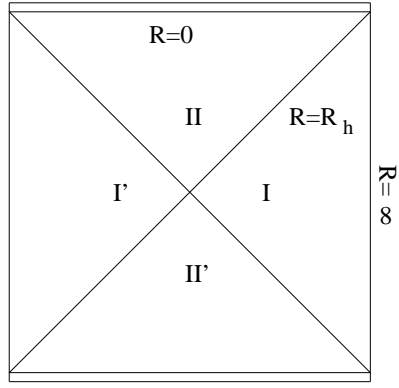}} 
\caption {{\small The Carter-Penrose Diagram for the $\omega=-3$ case.}} 
\label{omega_-3} 
\end{figure}

For $\omega=-3$, the general metric in Eq. (\ref{solutions}) and the 
$\phi$ field in Eq. (\ref{phisolutions_1}) reduce to the following 
\begin{eqnarray} 
ds^2 &=& - \left[ (a\,R)^2 - 2\sqrt{a\,R}\, M \right]\,dT^2 + 
\frac{dR^2}{(a\,R)^2 - 2\sqrt{a\,R}\,M}+R^2\,d\varphi^2\,, 
\label{metric_dilatonic}\\ 
e^{-2\phi} &=& \frac{1}{\sqrt{a\,R}}\,,\label{phi_dilatonic} 
\end{eqnarray} 
with $2M=b$ and $a=\frac{4\sqrt{3}}{3}|\lambda|$. 
In Fig. \ref{omega_-3} the Carter-Penrose diagram of the black hole 
solution for the case $\omega=-3$ is shown, which is analogous to 
the case $\omega=0$, where again the singularity at $R=0$ is a 
curvature singularity. 
Since we are now familiar with the whole formalism, we will be briefer in 
this section omitting several of the details. 

\subsection{Canonical formalism } 
With $\omega=-3$ the solution implies that 
$e^{-2\phi}=(a\,R)^{-\frac12}$. The value 
of $\omega$ also implies that $a=\frac{4\sqrt{3}}{3}|\lambda|$ 
(see Eq. (\ref{a_1})). 
The action then becomes, up to surface terms, 
\begin{eqnarray} \label{action_-3} 
S[\Lambda,\,R,\,\dot{\Lambda},\,\dot{R};\,N,\,N^r] &=& 
\int dt \int_0^{\infty} dr\, 
\left\{4\,(a\,R)^{-\frac12}N\lambda^2\Lambda R- 
(a\,R)^{-\frac12}N^{-1}\dot{\Lambda}\dot{R}+ 
\frac14 (a\,R)^{-\frac12} N^{-1} \Lambda R^{-1} \dot{R}^2\right. 
\nonumber \\ && + 
(a\,R)^{-\frac12} N^{-1} \dot{R} (N^r \Lambda)' + 
(a\,R)^{-\frac12} N^{-1}N^r\dot{\Lambda}R' 
-(a\,R)^{-\frac12} N^{-1}N^r(N^r)'\Lambda R' 
\nonumber \\ && 
-(a\,R)^{-\frac12} N^{-1}(N^r)^2 \Lambda' R' 
-\frac12 (a\,R)^{-\frac12} N^{-1}N^r \Lambda R^{-1}\dot{R}R' 
+\frac14 (a\,R)^{-\frac12} N^{-1}(N^r)^2\Lambda R^{-1}(R')^2 
\nonumber \\ && \left. 
+\frac14 (a\,R)^{-\frac12} N \Lambda^{-1}R^{-1}(R')^2 - 
(a\,R)^{-\frac12} N (\Lambda^{-1})'R' 
 -(a\,R)^{-\frac12}N\Lambda^{-1}R''\right\}\,. 
\end{eqnarray} 
Depending on the situation we use three different letters containing 
the same information but with slightly different numerical 
values. Thus, $a$, $\lambda$, and $l$ are related by 
$a^2=\frac{16}{3}\lambda^2 = \frac83 l^{-2}$, where $l$ is defined 
now as the AdS length.  From the action above, 
we obtain the conjugate momenta 
\begin{eqnarray} 
P_\Lambda &=& -N^{-1}(a\,R)^{-\frac12}\left(\dot{R}-R'N^r\right)\,,\\ 
P_R &=& -N^{-1}(a\,R)^{-\frac12} \left\{\dot{\Lambda}- 
(\Lambda N^r)'-\frac12 \Lambda R^{-1}(\dot{R}-R'N^r)\right\}\,. 
\end{eqnarray} 
By performing a Legendre transformation we obtain the Hamiltonian, 
which is a sum of constraints, i.e., 
\begin{eqnarray} \label{Hamiltonian_-3} 
\mathcal{H} &=& N\left\{ -(a\,R)^{\frac12}P_\Lambda P_R + 
\frac14 a^{\frac12}R^{-\frac12}\Lambda P_\Lambda^2 + 
(a\,R)^{-\frac12}\left[(\Lambda^{-1})'R'+ 
\Lambda^{-1}R''-\frac14 \Lambda^{-1}(R')^2R^{-1}\right]- 
4(a\,R)^{-\frac12}\lambda^2\Lambda R \right\} 
\nonumber \\ 
&& +\, N^r\left\{P_R R' + \Lambda (P_\Lambda)'\right\}\, 
\equiv\,NH+N^r H_r\,. 
\end{eqnarray} 
We can now write the action in Hamiltonian form, which reads 
\begin{equation} \label{haction_-3} 
S[\Lambda,\,R,\,P_{\Lambda},\,P_{R};\,N,\,N^r] = \int dt 
\int_0^{\infty} dr \left\{ P_{\Lambda}\dot{\Lambda}+P_R\dot{R}- 
NH-N^rH_r \right\}\,, 
\end{equation} 
with the constraints defined in Eq. (\ref{Hamiltonian_-3}). 
{}From here we derive the equations of motion for the canonical variables 
and respective canonical momenta 
\begin{eqnarray} \label{eom_-3_1} 
\dot{\Lambda} &=& -N (a\,R)^{\frac12} P_R + 
\frac12 N a^{\frac12}R^{-\frac12} 
\Lambda P_\Lambda + (N^r\Lambda)'\,,\\ \label{eom_-3_2} 
\dot{R} &=& -N (a\,R)^{\frac12} P_\Lambda + N^r R'\,, \\ \label{eom_-3_3} 
\dot{P}_\Lambda &=& -\frac14 N a^{\frac12}R^{-\frac12} P_\Lambda^2 - 
\left(N(a\,R)^{-\frac12}\right)'R'\Lambda^{-2} - 
\frac14 N (a\,R)^{-\frac12} (R')^2R^{-1}\Lambda^{-2} + 
4N(a\,R)^{-\frac12}\lambda^2R + N^rP_\Lambda'\,, \\ 
\dot{P}_R &=& \frac12 N a^{\frac12}R^{-\frac12} P_\Lambda P_R + 
\frac18 N a^{\frac12}R^{-\frac32} \Lambda P_\Lambda^2 - 
N a^{-\frac12} (\Lambda^{-1})'(R^{-\frac12})' \nonumber \\ 
&& + \frac14 N (a\,R)^{-\frac12} \Lambda^{-1} R^{-1} R'' 
- \left(\left(N a^{-\frac12}\right)'\Lambda^{-1} R^{-\frac12} 
\right)' 
- \frac12 N a^{-\frac12} \Lambda^{-1}(R^{-\frac12})'' \nonumber \\ 
\label{eom_-3_4} 
&& + 2 N a^{-\frac12} \lambda^2 \Lambda R^{-\frac12} + (N^rP_R)'\,. 
\end{eqnarray} 
For a correct variational principle to be applied, we have to find out 
what surface terms are left over from the variation performed, with the 
purpose of deriving the equations of motion in 
Eqs. (\ref{eom_-3_1})-(\ref{eom_-3_4}). These surface 
terms are, 
\begin{eqnarray} \label{surf_terms_-3} 
\textrm{Surface terms} &=& N (a\,R)^{-\frac12} R' \Lambda^{-2} 
\delta\Lambda -  N (a\,R)^{-\frac12} \Lambda^{-1} \delta R' + 
\left((a\,R)^{-\frac12}\right)'\Lambda^{-1} \delta R + 
\frac12(a\,R)^{-\frac12}\Lambda^{-1}R^{-1}R'\delta R \nonumber \\ 
&& - \left. N^rP_R\delta R + N^r\Lambda\delta P_\Lambda 
\right|_0^{\infty}\,. 
\end{eqnarray} 
In order to know the form of (\ref{surf_terms_-3}) for $r \to 0$, we 
assume 
\begin{eqnarray} \label{falloff_0_i_-3} 
\Lambda(t,r) &=& \Lambda_0+O(r^2)\,,\\ 
R(t,r)&=& R_0+R_2r^2+O(r^4)\,,\\ 
P_\Lambda(t,r) &=& O(r^3)\,\\ 
P_R(t,r) &=& O(r)\,,\\ 
N(t,r) &=& N_1(t)r+O(r^3)\,\\ 
N^r(t,r) &=& O(r^3)\,. \label{falloff_0_f_-3} 
\end{eqnarray} 
The surface terms become 
\begin{equation} \label{surf_terms_0_-3} 
\left. \textrm{Surface terms}\right|_{r=0} = - N_1(a\,R)^{-\frac12} 
\Lambda_0^{-1} 
\delta R_0\,. 
\end{equation} 
The asymptotic conditions for $r\rightarrow\infty$ are assumed as 
\begin{eqnarray} \label{falloff_inf_i_-3} 
\Lambda(t,r) &=& \frac{\sqrt{6}}{4}\,  l r^{-1} + 
2^{\frac54} l^{\frac52}\eta(t)r^{-\frac52} + O^\infty(r^{-3})\,,\\ 
R(t,r) &=& r + 2^{\frac34} l^{\frac32} 
\rho(t)r^{-\frac12}+O^\infty(r^{-1})\,,\\ 
P_\Lambda(t,r) &=& O^\infty(r^{-2})\,\\ 
P_R(t,r) &=& O^\infty(r^{-4})\,,\\ 
N(t,r) &=& R(t,r)'\Lambda(t,r)^{-1}(\tilde{N}_+(t)+O^\infty(r^{-5}))\,,\\ 
N^r(t,r) &=& O^\infty(r^{-2})\,, \label{falloff_inf_f_-3} 
\end{eqnarray} 
where $l$ is the AdS length. 
The surface terms, Eq. (\ref{surf_terms_-3}), for 
$r\rightarrow\infty$, are written as 
\begin{equation} \label{surf_terms_inf_-3} 
\left. \textrm{Surface terms}\right|_{r\rightarrow\infty} = 
\tilde{N}_+\delta M_+\,, 
\end{equation} 
where 
\begin{equation} \label{m_+_-3} 
M_+(t) = 2^53^{-\frac54}\,\eta(t)+2^23^{\frac12}\,\rho(t)\,. 
\end{equation} 
Therefore, the surface term added to (\ref{haction_-3}) is 
\begin{equation} \label{variationofsurfaceterm_-3} 
S_{\partial\Sigma}\left[\Lambda,R;N\right]=\int dt \left(2a^{-1} 
  (a\,R_0)^{\frac12}N_1 \Lambda_0^{-1}-\tilde{N}_+M_+\right)\,. 
\end{equation} 
With this surface term added we obtain a well defined variational 
principle. What remains after variation of the total action, 
Eq. (\ref{haction_-3}) and Eq. (\ref{variationofsurfaceterm_-3}), is 
\begin{equation} 
\int\,dt\,\left(2a^{-1}(a\,R_0)^{\frac12}\delta(N_1\Lambda_0^{-1}) 
-\delta\tilde{N}_+M_+\right)\,. 
\end{equation} 
The surface terms coming from the variation of the total action 
disappear as the result of the fixation of 
$\delta(N_1\Lambda_0^{-1})$ on the horizon, $r=0$, and of 
$\tilde{N}_+(t)$ at infinity, $r \to \infty$. 

\subsection{Reconstruction, canonical transformation, and action} 

Repeating the steps of the two previous sections, we now give the main 
results concerning $\omega=-3$. 
The metric function is given by the expression 
\begin{equation} 
F=(a\,R)^2 - 2\sqrt{a\,R}\,M\,. 
\label{metric_function_-3} 
\end{equation} 
The Killing time $T=T(t,r)$ is a function of $(t,r)$, where we find that 
\begin{equation} 
-T' = F^{-1} \Lambda P_\Lambda (a\,R)^{\frac12}\,, 
\label{tlinha_-3} 
\end{equation} 
which is equal to minus the conjugate momentum of the new variable $M$, 
or $P_M=P_M(t,r)\equiv -T'(t,r)$. 
With the help of Eq. (\ref{tlinha_-3}) we find $F=F(t,r)$ as a function 
of the canonical variables, 
\begin{equation} 
F=(R')^2\Lambda^{-2}-aRP_\Lambda^2\,. 
\end{equation} 
Summing up, the canonical transformations are 
\begin{eqnarray} \label{setcanonicaltrans_-3} 
R &=& \textrm{R}\,, \nonumber\\ 
M &=& \frac12 (a\,R)^{-\frac12} \left((a\,R)^2-F\right)\,, \nonumber \\ 
P_\textrm{R} &=& P_R - F^{-1}R^{-1}\Lambda^{-2}\left[RR'\Lambda 
  P_\Lambda' + \frac12 (R')^2 P_\Lambda \Lambda - RR''\Lambda 
  P_\Lambda + RR'\Lambda'P_\Lambda\right] \nonumber \\ 
&& +\frac14 R^{-1}\Lambda P_\Lambda - 4\lambda^2F^{-1}P_\Lambda 
R\Lambda\,, \nonumber \\ 
P_M &=& F^{-1} \Lambda P_\Lambda (a\,R)^{\frac12}\,. 
\end{eqnarray} 
In addition, the relevant inverse transformations back to the old variable 
$\Lambda$ and respective conjugate momentum $P_\Lambda$ are 
\begin{eqnarray} \label{inversetrans_l_-3} 
\Lambda &=& \left(F^{-1}(R')^2-FP_M^2\right)^{\frac12}\,, \\ 
P_\Lambda &=& (a\,R)^{-\frac12} FP_M 
\left(F^{-1}(R')^2-FP_M^2\right)^{-\frac12}\,. \label{inversetrans_pl_-3} 
\end{eqnarray} 
We have to show that this transformation is canonical. This requires 
using the identity 
\begin{eqnarray} \label{oddidentity_-3} 
P_\Lambda \delta\Lambda + P_R \delta R - P_M \delta M - 
P_{\textrm{R}}\delta\textrm{R} &=& \left( \frac12(a\,R)^{-\frac12}\delta 
R\ln 
  \left|\frac{(a\,R)^{-\frac12}R'+\Lambda P_\Lambda}{(a\,R)^{-\frac12}R'- 
\Lambda 
      P_\Lambda}\right|\right)'+ \nonumber \\ 
&& +\, \delta\left(\Lambda P_\Lambda + \frac12 (a\,R)^{-\frac12} R' \ln 
 \left|\frac{(a\,R)^{-\frac12} R'-\Lambda P_\Lambda}{(a\,R)^{-\frac12} R'+ 
\Lambda 
      P_\Lambda}\right|\right)\,. 
\end{eqnarray} 
We now integrate Eq. (\ref{oddidentity_-3}), in $r$, 
from $r=0$ to $r=\infty$. The first term on the right 
hand side of Eq. (\ref{oddidentity_-3}) vanishes due to the falloff 
conditions, Eqs. (\ref{falloff_0_i_-3})-(\ref{falloff_0_f_-3}) and 
Eqs. (\ref{falloff_inf_i_-3})-(\ref{falloff_inf_f_-3}). 
We obtain then the following expression 
\begin{eqnarray} \label{int_oddidentity_-3} 
\int_0^\infty\,dr\,\left(P_\Lambda \delta\Lambda + P_R \delta 
  R\right)-\int_0^\infty\,dr\,\left(P_M \delta M + 
P_{\textrm{R}}\delta\textrm{R} \right) &=& 
\delta\omega\,\left[\Lambda,\,R,\,P_\Lambda\right]\,, 
\end{eqnarray} 
where $\delta\omega\,\left[\Lambda,\,R,\,P_\Lambda\right]$ is a well 
defined functional, which is also an exact form. As above, this 
equality shows that the difference between the Liouville form of 
$\left\{R,\,\Lambda;\,P_R,\,P_\Lambda\right\}$ and the Liouville form 
of $\left\{\textrm{R},\,M;\,P_{\textrm{R}},\,P_M\right\}$ is an exact 
form, which implies that the set of transformations 
(\ref{setcanonicaltrans_-3}) is canonical. 

With this result, we write the asymptotic conditions of the new 
canonical coordinates for $r\rightarrow 0$ 
\begin{eqnarray} \label{newfalloff_0_i_-3} 
F(t,r) &=& 4 R_2^2 \Lambda_0^{-2} r^2 + O(r^4)\,, \\ 
\textrm{R}(t,r) &=& R_0+R_2\,r^2+O(r^4)\,, \\ 
M(t,r) &=& M_0 + M_2 \,r^2 + O(r^4)\,, \\ 
P_\textrm{R}(t,r) &=& O(r)\,, \\ 
P_M(t,r) &=& O(r)\,,\label{newfalloff_0_f_-3} 
\end{eqnarray} 
where we have 
\begin{eqnarray} \label{m_0_-3} 
M_0 &=& 2\,3^{-\frac34} l^{-2} R_0^2\left(\frac{\sqrt{2}}{2} l^{-1} 
R_0\right)^{-\frac12}\,, \\ 
M_2 &=& -l^{-2} R_0 R_2\left(\frac{\sqrt{2}}{2} l^{-1} 
  R_0\right)^{\frac12} + 
\frac14\left(\frac{\sqrt{2}}{2} l^{-1} R_0\right)^{-\frac12}3^{\frac14} 
\left[\frac{8}{3}l^{-2} R_0 R_2 - 4 R_2^2 \Lambda_0^{-2} 
\right]\,.\label{m_2_-3} 
\end{eqnarray} 
For $r\rightarrow \infty$, we have 
\begin{eqnarray} \label{newfalloff_inf_i_-3} 
F(t,r) &=& \frac{4}{3} l^{-2} \, r^2 
-\frac{16}{3}\left(\frac{\sqrt{2}\,l^{-1}}{6}\right)^{\frac12} 
\left(8\eta(t)+3^{\frac12}\rho(t)\right)\,r^{\frac12} 
+ O^{\infty}(r^{0})\,, \\ 
\textrm{R}(t,r) &=& r + 
\left(\frac{\sqrt{2}}{2}\,l^{-1}\right)^{-\frac32}\rho(t)\,r^{-\frac12}+ 
O^{\infty}(r^{-1})\,, \\ 
M(t,r) &=& M_+(t) + O^{\infty}(r^{-\frac12}) \,, \\ 
P_\textrm{R}(t,r) &=& O^\infty(r^{-4})\,, \\ 
P_M(t,r) &=& O^\infty(r^{-\frac95})\,, \label{newfalloff_inf_f_-3} 
\end{eqnarray} 
where $ M_+(t)$ is defined in Eq. (\ref{m_+_-3}). 

We now write the future constraint 
$M'$ as a function of the older constraints 
\begin{equation} 
M' = -\Lambda^{-1}\left( R' H +(a\,R)^{\frac12} P_\Lambda H_r \right)\,. 
\end{equation} 
Using the inverse transformations of $\Lambda$ and $P_\Lambda$ in 
Eqs. (\ref{inversetrans_l_-3}) and (\ref{inversetrans_pl_-3}), we 
obtain the same form for the old constraints as functions of the new 
variables 
\begin{eqnarray} \label{oldconstraintsinnewvariables_1_-3} 
H &=& - \frac{M'F^{-1}\textrm{R}'+F P_M 
P_{\textrm{R}}}{\left(F^{-1}(\textrm{R}')^2-F 
    P_M^2\right)^{\frac12}}\,, \\ 
H_r &=& P_M M' + P_{\textrm{R}} \textrm{R}'\,. 
\label{oldconstraintsinnewvariables_2_-3} 
\end{eqnarray} 
The new Hamiltonian, the total sum of the constraints, can now be 
written as 
\begin{equation} \label{newHamiltonian_-3} 
NH+N^rH_R = N^M M' + N^{\textrm{R}} P_{\textrm{R}}\,. 
\end{equation} 
The new multipliers are, using Eqs. 
(\ref{oldconstraintsinnewvariables_1_-3})-(\ref{newHamiltonian_-3}), 
\begin{eqnarray} \label{new_mult_1_-3} 
N^M &=& - \frac{N F^{-1} R'}{\left(F^{-1}(\textrm{R}')^2-F 
    P_M^2\right)^{\frac12}}+ N^r P_M \,, \\ 
N^{\textrm{R}} &=& - \frac{N F P_M}{\left(F^{-1}(\textrm{R}')^2-F 
    P_M^2\right)^{\frac12}}+ N^r R'\,. \label{new_mult_2_-3} 
\end{eqnarray} 
Using the inverse transformations 
Eqs. (\ref{inversetrans_l_-3})-(\ref{inversetrans_pl_-3}), and the 
identity $R=\textrm{R}$, we can write the new multipliers as functions 
of the old variables 
\begin{eqnarray} \label{mult_trans_1_-3} 
N^M &=& - NF^{-1}R'\Lambda^{-1}+ N^r F^{-1} \Lambda P_\Lambda 
(a\,R)^{\frac12} \,, \\ 
N^{\textrm{R}} &=& - N P_\Lambda (a\,R)^{\frac12} + N^r R'\,. 
\label{mult_trans_2_-3} 
\end{eqnarray} 
For $r\rightarrow 0$ we have, 
\begin{eqnarray} \label{mult_newfalloff_0_i_-3} 
N^M(t,r) &=& -\frac12  N_1(t) \Lambda_0 R_2^{-1} + O(r^2)\,,\\ 
N^{\textrm{R}}(t,r) &=& O(r^4)\,,\label{mult_newfalloff_0_f_-3} 
\end{eqnarray} 
and for $r\rightarrow\infty$ we have 
\begin{eqnarray} \label{mult_newfalloff_inf_i_-3} 
N^M(t,r) &=& -\tilde{N}_+(t) +  O^\infty(r^{-2})\,,\\ 
N^{\textrm{R}}(t,r) &=& O^\infty(r^{-\frac12})\,. 
\label{mult_newfalloff_inf_f_-3} 
\end{eqnarray} 
Again, for $r\rightarrow0$, fixing $N^M(t,r)$, which means fixing 
$N_1(t) \Lambda_0 R_2^{-1}$, is not 
equivalent to fixing $N_1 \Lambda_0^{-1}$. 
It is thus necessary to rewrite $N^M$ for $r\rightarrow0$. 
So, assuming $M_0$ as a function of $R_0$ allows one to define the 
horizon radius 
$R_0\equiv R_{\textrm{h}}(M_0)$. We are thus working in the 
domain where $M_0>0$, the 
domain of the classical black hole solution. 
The variation of $R_0$ 
is given in terms of the variation of $M_0$ in the expression 
\begin{equation} \label{deltar_0_-3} 
\delta R_0 = \frac43 a^{-1} (a\,R)^{-\frac12}\delta M_0\,. 
\end{equation} 
Remember that $a=\frac{4\sqrt{3}}{3}|\lambda|$. 
The new multiplier $\tilde{N}^M$ is obtained from the old $N^M$ as 
\begin{equation} \label{new_n_m_-3} 
\tilde{N}^M = - N^M 
\left[(1-g) + \frac12 g l^2 R_0^{-1}\right]^{-1}\,, 
\end{equation} 
where $l^2=2\lambda^2$, and $g(r)=1+O(r^2)$ for $r\rightarrow0$ and 
$g(r)=O^\infty(r^{-5})$ for $r\rightarrow\infty$. 
This new multiplier, function of the old multiplier, $\tilde{N}^M$, has 
as its properties for $r\rightarrow\infty$ 
\begin{equation} 
\tilde{N}^M(t,r) = \tilde{N}_+(t) + O^\infty(r^{-2})\,, 
\end{equation} 
and as its properties for $r\rightarrow0$ 
\begin{equation} 
\tilde{N}^M(t,r) = \tilde{N}_0^M(t) + O(r^{2})\,, 
\end{equation} 
where $\tilde{N}_0^M$ is given by 
\begin{equation} 
\tilde{N}_0^M = 2\lambda^2 N_1 R_0 R_2^{-1} \Lambda_0\,. 
\end{equation} 
When the constraint $M'=0$ holds, the last expression is 
\begin{equation} 
\tilde{N}_0^M =  N_1 \Lambda_0^{-1} \,. 
\end{equation} 
With this new constraint $\tilde{N}^M$, fixing $N_1 \Lambda_0^{-1}$ 
or fixing $\tilde{N}^M$ is equivalent, there being no problem with 
$N^{\textrm{R}}$, which is left as determined in Eq. 
(\ref{mult_newfalloff_inf_f_-3}). 

The new action is then, summing both the bulk and the surface terms, 
\begin{eqnarray} 
S\left[M, \textrm{R}, P_M, P_{\textrm{R}}; \tilde{N}^M, 
  N^{\textrm{R}}\right] &=& \int \,dt\, \int_0^\infty \, dr \, 
\left( P_M\dot{M} + P_\textrm{R} \dot{\textrm{R}} - 
N^{\textrm{R}}P_{\textrm{R}} + \tilde{N}^M 
\left[(1-g) + \frac12 g\,l^{2}\,R_0^{-1}\right]\,M'\right)+ 
\nonumber \\ 
&&  \int \, dt \, \left(2a^{-1} (a\,R_0)^{\frac12} \tilde{N}_0^M - 
  \tilde{N}_+ M_+ \right)\,. \label{newaction_-3} 
\end{eqnarray} 
The new equations of motion are now 
\begin{eqnarray} \label{new_eom_1_-3} 
\dot{M} &=& 0\,, \\ 
\dot{\textrm{R}} &=& N^{\textrm{R}}\,, \\ 
\dot{P}_M &=& (N^M)'\,, \\ 
\dot{P}_{\textrm{R}} &=& 0\,, \\ 
M' &=& 0\,, \\ 
P_{\textrm{R}} &=& 0\,. \label{new_eom_6_-3} 
\end{eqnarray} 
Here we understood $N^M$ to be a function of the new constraint, 
defined through Eq. (\ref{new_n_m_-3}). The resulting boundary terms 
of the variation of this new action, Eq. (\ref{newaction_-3}), are, first, 
terms proportional to $\delta M$ and $\delta \textrm{R}$ on the 
initial and final hypersurfaces, and, second, the term 
$\int \, dt \, \left(2a^{-1} (a\,R_0)^{\frac12} \delta\tilde{N}_0^M - 
M_+ \delta\tilde{N}_+ \right)$. 
Here we used the expression in Eq. (\ref{deltar_0_-3}). The action in 
Eq. (\ref{newaction_-3}) yields the equations of motion, Eqs. 
(\ref{new_eom_1_-3})-(\ref{new_eom_6_-3}), provided that we fix 
the initial and final values of the new canonical variables and that 
we also fix the values of $\tilde{N}^M_0$ and of $\tilde{N}_+$. Thanks to 
the redefinition of the Lagrange multiplier, from $N^M$ to 
$\tilde{N}^M$, the fixation of those quantities, $\tilde{N}^M_0$ and 
$\tilde{N}_+$, has the same meaning it had before the 
canonical transformations and the redefinition of $N^M$. 
This same meaning is guaranteed through the use of our gauge freedom to 
choose the multipliers, and at the same time not fixing the boundary 
variations independently of the choice of Lagrange multipliers, which 
in turn allow us to have a well defined variational principle for the 
action. 

\subsection{Hamiltonian reduction} \label{hr_-3} 

We now solve the constraints in order to reduce to the true dynamical 
degrees of freedom. The equations of motion, Eqs. 
(\ref{new_eom_1_-3})-(\ref{new_eom_6_-3}), allow us to write $M$ as an 
independent function of the radial coordinate $r$, 
\begin{equation} 
M(t,r)=\textbf{m}(t)\,. 
\label{m_-3} 
\end{equation} 
The reduced action, with the constraints and Eq (\ref{m_-3}) taken into 
account, is 
\begin{equation} \label{red_action_-3} 
S 
\left[\textbf{m},\textbf{p}_{\textbf{m}};\tilde{N}_0^M,\tilde{N}_+\right] 
= \int 
dt\,\textbf{p}_{\textbf{m}} \dot{\bf{m}}-\textbf{h}\,, 
\end{equation} 
where 
\begin{equation} \label{new_p_m_-3} 
\textbf{p}_{\textbf{m}} = \int_0^\infty dr\,P_M\,, 
\end{equation} 
and the reduced Hamiltonian, $\textbf{h}$, is now written as 
\begin{equation} \label{red_Hamiltonian_-3} 
\textbf{h}(\textbf{m};t) = -2a^{-1} 
(a\,R_{\textrm{h}})^{\frac12} \tilde{N}_0^M + 
\tilde{N}_+ \textbf{m}\,, 
\end{equation} 
with $R_{\textrm{h}}$ being the horizon 
radius. We also have that $\textbf{m}>0$. 
The equations of motion are then 
\begin{eqnarray} \label{red_eom_1_-3} 
\dot{\textbf{m}} &=& 0\,, \\ 
\dot{\textbf{p}}_{\textbf{m}} &=& \frac43 a^{-2} (R_{\textrm{h}})^{-1} 
\tilde{N}_0^M - \tilde{N}_+\,. 
\label{red_eom_2_-3} 
\end{eqnarray} 
Here $\textbf{m}$ is equal to the mass parameter $M$ of the 
classical solution in Eq. (\ref{metric_dilatonic}). 
The second equation of motion, Eq. (\ref{red_eom_2_-3}), 
describes the time evolution of the 
difference of the Killing times on the horizon and at infinity, due to 
$ 
\textbf{p}_{\textbf{m}} = T_0 - T_+\, 
$ 
and Eq. (\ref{new_p_m_-3}). 

\subsection{Quantum theory and partition function} 

The steps developed in Sections \ref{btz} and \ref{zero} 
can be readily used here. So we do nont spell out the 
corresponding calculations in detail. 

\subsection{Thermodynamics} 

We can now build the partition function for this system, with the 
Hamiltonian 
given in Eq. (\ref{red_Hamiltonian_-3}). The steps are the same as was the 
case 
with $\omega=\infty$ and $\omega=0$. 
The thermodynamic ensemble is also the canonical ensemble. 
Thus, the operator is 
\begin{equation} 
K\left(\textbf{m};\mathcal{T},\Theta\right) = \exp 
\left[-i\textbf{m}\mathcal{T}+ 2\,i\,a^{-1}\, 
(a\,R_{\textrm{h}})^{\frac12}\Theta 
\right]\,. 
\end{equation} 
Again, with $\mathcal{T}=-i\beta$ and $\Theta=-2\pi i$, we write the 
general form 
of the partition function as 
\begin{equation} 
\mathcal{Z} = Tr\left[ \hat{K}(-i\beta, -2\pi i) \right]\,. 
\end{equation} 
This is realized as 
\begin{equation} \label{partition_function_ren_-3} 
\mathcal{Z}_{\textrm{ren}}(\beta) = \mathcal{N} \int_{A'} 
\widetilde{\mu}\,dR_{\textrm{h}}\,\exp(-I_*)\,, 
\end{equation} 
where $\mathcal{N}$ is a normalization factor, 
$A'$ is the domain of integration, and the function 
$I_*(R_{\textrm{h}})$, a kind of an 
effective action \cite{york1}. $I_*(R_{\textrm{h}})$, is written as 
\begin{equation} \label{eff_action_-3} 
I_*(R_{\textrm{h}}):= \frac{\beta}{2} (a\,R_{\textrm{h}})^{\frac32} 
-4\pi a^{-1} (a\,R_{\textrm{h}})^{\frac12}\,. 
\end{equation} 
The domain of integration, given by $A'$, is defined by the inequality 
$R_{\textrm{h}}\geq 0$. 
The new weight factor $\widetilde{\mu}$ includes the Jacobian of the 
change of variables, which amounts to $\partial \textbf{m} /\partial 
R_{\textrm{h}}$. 
Its critical point is at $R_\textrm{h}^+=\frac83 a^{-2} \pi\beta^{-1}$. 
The action evaluated at the critical point is 
\begin{equation} 
I_*(R_{\textrm{h}}^+) = -a^{-\frac32} \beta^{-\frac12} 
\left(\frac83 \pi\right)^\frac32 \,. 
\label{action_at_critical_point_-3} 
\end{equation} 
{}From Eq. (\ref{action_at_critical_point_-3}) one sees that 
$I_*(R_{\textrm{h}}^+)<0$. 
It is seen that the critical point is a minimum. 
This implies that 
\begin{eqnarray} 
\mathcal{Z}_{\textrm{ren}}(\beta) &\approx& 
P\exp\left[-I_*(R_{\textrm{h}}^+)\right]\,\nonumber\\ 
&=& P\exp\left[-\frac{\beta}{2}(a\,R_{\textrm{h}}^+)^\frac32\right]\,. 
\end{eqnarray} 
We can now derive the basic thermodynamic results, as long as 
the approximation for the saddle point is valid, which means that we 
have to work in the neighborhood of the classical solution 
(\ref{metric_dilatonic}), where the critical point dominates. 
The expected value of the energy $E$ is 
\begin{equation} 
\left\langle E \right\rangle = -\frac{\partial}{\partial\beta}\ln 
\mathcal{Z}_{\textrm{ren}}(\beta) \approx \frac12 
(a\,R_{\textrm{h}}^+)^\frac32 = \textbf{m}^+\,. 
\end{equation} 
Here we have 
\begin{eqnarray} 
\textbf{T}&=&\left(\frac{3^\frac34\left|\lambda\right|^\frac32
\textbf{m}^+}{2^\frac12\pi^\frac32}\right)^\frac23\,, 
\label{temperature_-3}\\ 
\textbf{m}^+(\beta)&=&\frac12 a^{-\frac32} 
\beta^{-\frac32} \left(\frac83\pi\right)^{\frac32}\,. 
\label{mass_-3} 
\end{eqnarray} 
The derivative of 
$\textbf{m}^+(\beta)$ with respect to $\beta$ is negative, 
which means that the heat capacity is positive. 
The system is thus stable. Finally the entropy is given by 
\begin{equation} 
S = 
\left(1-\beta\frac{\partial}{\partial 
\beta}\right)(\ln\mathcal{Z}_{\textrm{ren}}(\beta)) 
\approx 4 \pi \sqrt{a^{-1}R_{\textrm{h}}^+}\,. 
\end{equation} 
The result recovers the entropy for the three-dimensional dilatonic 
black hole with $\omega=-3$ (see also \cite{ads3_bh}). 

\section{Conclusions} 
\label{conclusions} 

We have continued the Louko-Whiting program of studying, through
Hamiltonian methods, the thermodynamic properties of black holes in
several theories in different dimensions (see \cite{louko1} and
\cite{louko2}-\cite{kunstatter2}).

Specifically we have calculated the thermodynamic properties of black
hole solutions with asymptotic infinities that allow a well formulated
Hamiltonian formalism in three-dimensional Brans-Dicke
dilaton-gravity. Only certain values of the Brans-Dicke parameter
$\omega$ are allowed in this juncture.  The corresponding theories are
general relativity, i.e., $\omega\to\infty$, a dimensionally reduced
cylindrical four-dimensional general relativity theory, i.e.,
$\omega=0$, and a theory representing a class of dilaton-gravity
theories, with a typical $\omega$ given by $\omega=-3$.  Within a
three-dimensional context, we have built a framework for the classical
Hamiltonian theory, where the metric functions are used as canonical
variables, and where one foliates the spacetime with equal time
spacelike hypersurfaces. These hypersurfaces go from the bifurcation
circle of the horizon on the left, to the asymptotic anti-de Sitter
infinity on the right.  Then we have performed a canonical
transformation based on a reconstruction of the canonical variables.
This canonical transformation is an adaptation of Kucha\v{r}'s work
\cite{kuchar} to the Louko-Whiting method
\cite{louko1,louko2,louko3,louko4,louko5}.  In \cite{kuchar} the
boundaries are the left and right infinities of the Kruskal diagram
(or if one wishes of the Carter-Penrose diagram),
whereas in \cite{louko1,louko2,louko3,louko4,louko5} and here, with a
thermodynamic goal in mind, the boundaries are the bifurcation sphere,
here a 1-sphere or circle, and the right infinity.  When the classical
equations hold one finds that the new canonical coordinate is indeed
the physical parameter of the classical solutions of the dilatonic
black holes in three dimensions, i.e., the black hole mass $M$. Its
conjugate momentum, $P_M=-T'$, is the spatial derivative of the
Killing time.  It survives the reduction, i.e., the elimination of
constraints, becoming at last the difference between the Killing time
at the bifurcation circle and at infinity. Its equation of motion is
the equation for the evolution of the difference between Killing
times, with respect to the time parameter $t$ of each
hypersurface. The other coordinate, $\textrm{R}$, and respective
momentum, $P_\textrm{R}$, vanish after reduction, being pure gauge.
With the new variables, come new constraints and new Lagrange
multipliers. We then reduce the Hamiltonian to an unconstrained theory
with one pair of canonical coordinates.  Then one performs a
quantization of the theory, where one replaces the functions of the
physical observables for operators of a Hilbert space.  Constructing
then the Schr\"odinger evolution operator from the Hamiltonian and
taking the trace on a suitable basis, we obtain the partition function
of the canonical thermodynamic ensemble. This ensemble is well defined
and under suitable conditions the classical Euclidean solutions
dominate the partition function, yielding the thermodynamics of the
systems.

To sum up, as noted above, this formalism was previously applied to
two \cite{louko2} (see also \cite{kunstatter1,kunstatter2}), four
\cite{louko1,louko3,louko5,bose}, and five dimensions \cite{louko4} in
several different theories.  Here we have applied to three dimensions
in a quite general dilaton-gravity theory.  We have shown that in
three-dimensional theories with well defined asymptotics the formalism
fits well.  As in other instances, the negative cosmological constant
has had a stabilizing role to play here in what the thermodynamic
results are concerned.  Notwithstanding, several modifications were
needed. First, in the powers of the fall-off conditions, and second
due to the presence of the scalar dilaton field, which was reflected
in the fact that it changed the powers of the radial $R$ coordinate,
to name a few.  Although, in order to build a three-dimensional
Lorentzian Hamiltonian theory, these modifications had to be made, we
have derived a quantum theory and a statistical description of the
systems in question, and found the corresponding thermodynamics, with
precise values for the temperature and entropy of the black holes
studied.

\section*{Acknowledgments} 

GASD is supported by grant SFRH/BD/2003 from FCT.  This work was 
partially funded by Funda\c c\~ao para a Ci\^encia e Tecnologia (FCT) 
- Portugal, through project POCI/FP/63943/2005. 


\end{document}